\documentclass[aps, 12pt, amsmath, amssymb, showpacs, preprint]{revtex4}
\usepackage{amsmath}
\usepackage{amssymb}
\usepackage{textcomp}
\usepackage{graphicx}% Include figure files
\usepackage{bm}% bold math
\usepackage{color}

\newcommand{\be}{\begin{equation}}
\newcommand{\ee}{\end{equation}}
\newcommand{\ba}{\begin{eqnarray}}
\newcommand{\ea}{\end{eqnarray}}
\newcommand{\nn}{\nonumber\\&&}
\newcommand{\ns}{\nonumber\\}
\newcommand{\dmass}{\frac{\partial}{\partial m}}
\newcommand{\ddmass}{\partial_m}
\newcommand{\dm}{\frac{\partial}{\partial \mu}}
\newcommand{\dn}{\frac{\partial}{\partial n}}
\newcommand{\gm}{\gamma^{\mu}}
\newcommand{\dmu}{\partial_{\mu}}
\newcommand{\dnu}{\partial_{\nu}}

\newcommand{\dmuu}{\partial^{\mu}}
\newcommand{\dnuu}{\partial^{\nu}}
\newcommand{\dz}{\partial_{z}}
\newcommand{\du}{\partial_{u}}

\begin{document}
\title{Low-energy theorems and spectral density of the Dirac operator in AdS/QCD}

\author{P.N. Kopnin}
\email{kopnin@itep.ru} \affiliation{ITEP, Moscow}
\affiliation{MIPT, Moscow}

\preprint{ITEP-TH-30/09}

%\date{\today}
\begin{abstract}

We study the low-energy theorems of QCD from the point of view of
the dual AdS/QCD models and demonstrate that these models are
compatible with the theorems in the chiral limit, i.e. the arising
expressions have the same analytical behavior at the pole when the
quark mass tends to zero. Low-energy theorems are formulated in
terms of the spectral density of the Dirac operator. In order to
calculate the spectral density in the dual holographic models we
express it in terms of a partition function of a QCD-like theory.

\end{abstract}
\pacs{11.25.Tq, 12.38.Lg, 12.39.Fe}

\nopagebreak[4]

\maketitle

%\tableofcontents

\nopagebreak[4]

%%%%%%%%%%%%%%%%%%%%%%%%%%%%%%%%%%%%%%%%%%%%%%%%%%%%%%%%%%%%%%%%%%%%%%%%%%%%%%%%%%%%%%%%%%%%%%%%%%%%%%%%%
%%%%%%%%%%%%%%%%%%%%%%%%%%%%%%%%%%%%%%%%%%%%%%%%%%%%%%%%%%%%%%%%%%%%%%%%%%%%%%%%%%%%%%%%%%%%%%%%%%%%%%%%%
%%%%%%%%%%%%%%%%%%%%%%%%%%%%%%%%%%%%%%%%%%%%%%%%%%%%%%%%%%%%%%%%%%%%%%%%%%%%%%%%%%%%%%%%%%%%%%%%%%%%%%%%%
%%%%%%%%%%%%%%%%%%%%%%%%%%%%%%%%%%%%%%%%%%%%%%%%%%%%%%%%%%%%%%%%%%%%%%%%%%%%%%%%%%%%%%%%%%%%%%%%%%%%%%%%%

\section{Introduction}

In the recent past new approaches to studying the low-energy
dynamics of QCD based on the idea of AdS/CFT correspondence have
emerged. The idea of this correspondence was first put forward in
\cite{adscft}. The prescription of the latter consists in using a
classical multidimensional string theory to describe a quantum
four-dimensional field theory in a strong coupling regime. Namely,
one puts into correspondence operators in the quantum field theory
and multidimensional classical fields on the string theory side
using their transformation properties with respect to global
symmetries of the field theory and isometries of the string theory
background space-time. Then the  AdS/CFT conjecture states that
the generating functional of the quantum field theory equals the
exponent of the action of the gravity theory in which the
multidimensional fields are confined to classical trajectories and
their boundary values are set to be equal to the sources of the
corresponding operators on the four-dimensional side:

\be \mathcal{Z}_{QFT}[J_i(x_{\mu})] = \left.\exp(i S_{Gravity\
Cl.})\right|_{\Phi_i(x_{\mu},y_M=0)=J_i(x_{\mu})}.\label{prescription}
\ee

This correspondence has been formulated precisely in the case of a
$\mathcal{N}=4$ Super Yang-Mills theory on one side and type IIB
string theory in the $AdS_5\times S^5$. The coupling $g_{YM}$ and
the number of colors $N_c$ of the field theory are related to the
radius of the sphere $S^5$ and $AdS_5$, string tension
$\alpha'^{-1}$ and coupling $g_{string}$ as well as to the 't
Hooft constant $\lambda'$:

\be \frac{R^4}{\alpha'^2}=4\pi g_{string} N_c=g^2_{YM}
N_c=\lambda'.\label{'thooftconst}\ee

When we move from the $\mathcal{N}=4$ Super Yang-Mills theory
towards the real-life QCD decreasing the number of supersymmetries
and destroying the initial conformal symmetry we start facing more
and more complex theories on the gravity side of the
correspondence. In order to explore QCD from the holographic point
of view two basically different approaches have been devised,
so-called "top-down" and "bottom-up". In the former, one starts
with a string theory setup and works all the way down to warp the
ten-dimensional geometry so that it reflects the dynamics of QCD
on its four-dimensional boundary, e.g. \cite{sakaisugimoto1},
\cite{karchkatz}. In the latter one formulates a minimal model
that holographically reproduces the symmetries and dynamics of
QCD. Such models were proposed in \cite{hardwall},
\cite{softwall1}, \cite{daroldpomarol}. One of their advantages is
that these simple five-dimensional toy-models allow us to study
the basic properties of QCD without the complexity of the
"top-down" approach. For instance, masses, decay rates and
couplings of the lightest mesons as well as the
Gell-Mann--Oakes--Renner relationship for the pion mass have been
studied in \cite{hardwall}, the linear confinement -- in
\cite{softwall1} and the chiral symmetry breaking -- in
\cite{krikun}.

However simple those models may be, they need to reproduce a set
of exact equations that describe the low-energy dynamics of QCD --
the so-called low-energy theorems. These theorems are derived from
axial Ward identities in QCD and are formulated for two-, three-
and four-point correlators. Their complete list can be found in
\cite{leutwyler}. We will focus on the following theorems in the
$N_f=2$ case \cite{gorsky}, \cite{smilga}:

\ba \frac{i}{V}\int d^4x d^4y \left< \delta_{ij}S_0(x)S_0(y) -
P_i(x)P_j(y) \right> &=& -\frac{G^2_{\pi}\delta_{ij}}{m^2_{\pi}} +
\delta_{ij}\frac{B^2}{8\pi^2}(L_3-4L_4+3)\ns &=& 2\delta_{ij} \int
d\lambda \left( \frac{m \dmass \rho(\lambda,m)}{(\lambda^2+m^2)} -
\frac{2m^2 \rho(\lambda,m)}{(\lambda^2+m^2)^2} \right),
\label{sspp}\\
\frac{i}{V}\int d^4x d^4y \left< S_i(x)S_j(y) -
\delta_{ij}P_0(x)P_0(y) \right> &=& -8\delta_{ij}B^2L_7  =
\delta_{ij}\int d\lambda \frac{4m^2
\rho(\lambda,m)}{(\lambda^2+m^2)^2}\ns &-& 2\delta_{ij} \frac{\int
d^4x d^4y \left< Q(x)Q(y) \right> }{m^2V},
\label{ppss}\\
\frac{i}{V}\int d^4x d^4y \left< P_3(x)P_0(y) \right> &=&
-\frac{G_{\pi}\tilde{G}_{\pi}}{m^2_{\pi}} = 2(m_u-m_d)m \int
d\lambda \frac{\rho(\lambda,m)}{(\lambda^2+m^2)^2} \ns &-&
(m_u-m_d)\frac{\int d^4x d^4y \left< Q(x)Q(y) \right>
}{m^3V}.\label{pp} \ea

Here $i,j$ are the adjoint flavor indices of the scalar and
pseudoscalar currents $S(x)$ and $P(x)$; $m_{\pi}$ is the pion
mass; $G_{\pi}=2F_{\pi}B=\dfrac{F_{\pi}m^2_{\pi}}{m}$ is its
pseudoscalar decay constant;
$\tilde{G}_{\pi}=4(m_d-m_u)\dfrac{B^2L_7}{F_{\pi}}$; $Q(x) =
\dfrac{g^2\theta}{32\pi^2} \mathrm{tr}
F_{\mu\nu}(x)\tilde{F}^{\mu\nu}(x)$ is the topological charge
density; $m$ is the quark mass, or, in the case of different quark
masses, $m=\dfrac{1}{2}(m_u+m_d)$; $V$ is the Euclidean
four-volume; $\rho(\lambda,m)$ is the spectral density of the
Dirac operator, and $L_i$ are the constants of the next-to-leading
order effective Chiral Lagrangian (see, e.g.,
\cite{leutwyler,handbook}).

The aim of this paper is to demonstrate that these theorems are
satisfied in the dual AdS/QCD models in the chiral limit, and
therefore we will focus on the leading terms in the expansion in
powers of quark mass $m$ of the relevant expressions. We leave the
case of a finite mass to future studies.

In the next section of the paper {\bf \ref{2point}} we compute the
necessary two-point correlators from the AdS/QCD point of view, in
the section {\bf \ref{chirallag}} we discuss the implications of
the AdS/QCD models for the Chiral Lagrangian, and in the section
{\bf \ref{spectrum}} we present a way to calculate the
aforementioned spectral density in the AdS/QCD framework. Finally,
in the section {\bf \ref{compatibility}} we discuss the
compatibility of the low-energy theorems with the AdS/QCD models.

%%%%%%%%%%%%%%%%%%%%%%%%%%%%%%%%%%%%%%%%%%%%%%%%%%%%%%%%%%%%%%%%%%%%%%%%%%%%%%%%%%%%%%%%%%%%%%%%%%%%%%%%%
%%%%%%%%%%%%%%%%%%%%%%%%%%%%%%%%%%%%%%%%%%%%%%%%%%%%%%%%%%%%%%%%%%%%%%%%%%%%%%%%%%%%%%%%%%%%%%%%%%%%%%%%%
%%%%%%%%%%%%%%%%%%%%%%%%%%%%%%%%%%%%%%%%%%%%%%%%%%%%%%%%%%%%%%%%%%%%%%%%%%%%%%%%%%%%%%%%%%%%%%%%%%%%%%%%%
%%%%%%%%%%%%%%%%%%%%%%%%%%%%%%%%%%%%%%%%%%%%%%%%%%%%%%%%%%%%%%%%%%%%%%%%%%%%%%%%%%%%%%%%%%%%%%%%%%%%%%%%%

\subsection{Action and fields\label{action}}

We will consider the following five-dimensional action of the
AdS/QCD models \cite{hardwall}, \cite{softwall1}, \cite{softwall}:

\be S_{5D} = \int d^5x \sqrt{g}e^{-\Phi} \textrm{tr}\left\{
\Lambda^2 \left( |D X|^2+\frac{3}{R^2}|X|^2
+\frac{\kappa}{R^2}|X|^4\right) -\frac{1}{4g^2_5}(F_L^2+F_R^2)
\right\}.\label{full5daction}\ee with a metric $ds^2 =
\dfrac{R^2}{z^2}(-dz^2+dx_{\mu}dx^{\mu})\equiv g_{MN}dx^M dx^N
\equiv \dfrac{R^2}{z^2}\eta_{MN}dx^M dx^N$.

Here $g\equiv det(g_{MN})$, $\Phi$ is a dilaton whose profile
depends on the choice of a particular model. In the so-called
hard-wall model \cite{hardwall} $\Phi(z)\equiv 0$ and the bulk
five-dimensional space has a boundary at $z=z_m$ where a uniform
Neumann boundary condition is imposed. In the soft-wall models
\cite{softwall1}, \cite{softwall} the bulk space stretches over
the whole range $0 \leqq z<\infty$ and the dilaton profile is set
to be asymptotically parabolic: \be \Phi(z)\sim \lambda z^2,
z\rightarrow\infty \label{dilaton}\ee so that the linear Regge
trajectory is reproduced.

We introduce two gauge fields $L^a_{\mu}$ and $R^a_{\mu}$ in the
adjoint representation of the $SU_L(N_f)$ and $SU_R(N_f)$ gauge
groups respectively with curvatures $F_L=dL-iL\wedge L,\
F_R=dR-iR\wedge R$, and a scalar field $X^{\alpha\beta}$ with mass
$-3/R^2$ in the bifundamental representation of $SU_L(N_f)\times
SU_R(N_f)$ that interacts with the gauge fields: $DX=dX-iLX+iXR$.

According to the AdS/CFT prescription (\ref{prescription}) the
fields on the AdS boundary act as sources of the QCD currents:

\ba L^a_{\mu}(x,z=0)&=&\mbox{source of
}\bar{q}_L(x)\gamma_{\mu}t^aq_L(x),\ns
R^a_{\mu}(x,z=0)&=&\mbox{source of
}\bar{q}_R(x)\gamma_{\mu}t^aq_R(x),\ns
\lim_{z\rightarrow0}\frac{2}{z}X^{\alpha\beta}(x,z)&=&\mbox{source
of }\bar{q}^{\alpha}_L(x)q^{\beta}_R(x).\label{fields}\ea

In the AdS action $g_5$ is the 5D coupling constant defined by
means of a comparison of the vector two-point correlator with the
QCD sum rules \cite{hardwall}, \cite{sumrules}:
$\dfrac{g_5^2}{R}=\dfrac{12\pi^2}{N_c}$; $\Lambda$ is the
normalization factor for the scalar field which is fixed by
comparing the pseudoscalar two-point correlator in the large
momentum regime with the sum rules \cite{sumrules}, \cite{krikun}:
$\Lambda^2 R^3 = \dfrac{N_c}{4\pi^2}$. Hard-wall models generally
do not include the effective potential of the scalar field
$V(X)=\dfrac{\kappa}{R^2}|X|^4$ although it is necessary in the
soft-wall models \cite{softwall}.

%%%%%%%%%%%%%%%%%%%%%%%%%%%%%%%%%%%%%%%%%%%%%%%%%%%%%%%%%%%%%%%%%%%%%%%%%%%%%%%%%%%%%%%%%%%%%%%%%%%%%%%%%
%%%%%%%%%%%%%%%%%%%%%%%%%%%%%%%%%%%%%%%%%%%%%%%%%%%%%%%%%%%%%%%%%%%%%%%%%%%%%%%%%%%%%%%%%%%%%%%%%%%%%%%%%
%%%%%%%%%%%%%%%%%%%%%%%%%%%%%%%%%%%%%%%%%%%%%%%%%%%%%%%%%%%%%%%%%%%%%%%%%%%%%%%%%%%%%%%%%%%%%%%%%%%%%%%%%
%%%%%%%%%%%%%%%%%%%%%%%%%%%%%%%%%%%%%%%%%%%%%%%%%%%%%%%%%%%%%%%%%%%%%%%%%%%%%%%%%%%%%%%%%%%%%%%%%%%%%%%%%

\section{Pseudoscalar and scalar two-point correlators\label{2point}}

We shall first discuss the simpler case of pseudoscalar currents.

%%%%%%%%%%%%%%%%%%%%%%%%%%%%%%%%%%%%%%%%%%%%%%%%%%%%%%%%%%%%%%%%%%%%%%%%%%%%%%%%%%%%%%%%%%%%%%%%%%%%%%%%%
%%%%%%%%%%%%%%%%%%%%%%%%%%%%%%%%%%%%%%%%%%%%%%%%%%%%%%%%%%%%%%%%%%%%%%%%%%%%%%%%%%%%%%%%%%%%%%%%%%%%%%%%%
%%%%%%%%%%%%%%%%%%%%%%%%%%%%%%%%%%%%%%%%%%%%%%%%%%%%%%%%%%%%%%%%%%%%%%%%%%%%%%%%%%%%%%%%%%%%%%%%%%%%%%%%%
%%%%%%%%%%%%%%%%%%%%%%%%%%%%%%%%%%%%%%%%%%%%%%%%%%%%%%%%%%%%%%%%%%%%%%%%%%%%%%%%%%%%%%%%%%%%%%%%%%%%%%%%%

\subsection{Pseudoscalar currents\label{pseudoscalar}}

As compared to the scalar five-dimensional field, the pseudoscalar
field is less sensitive to the shape of the effective potential
$V(X)$. For that reason we will use a simpler hard-wall model
\cite{hardwall}, \cite{krikun} while calculating the
$\left<PP\right>$ correlator. We will need the following part of
the 5D action: \be S_{5D} = \int d^5x \sqrt{g}\Lambda^2
\textrm{tr} \left( |\partial X|^2+\frac{3}{R^2}|X|^2
\right).\label{5dlag}\ee

According to the AdS/QCD prescription (\ref{fields}),
$\left.\dfrac{2}{z}X^{\alpha\beta}(z,x)\right|_{z\rightarrow 0}
\leftrightarrow \bar{q}_L^{\alpha}(x) q_R^{\beta}(x)$. This means
that the pseudoscalar current corresponds to
$i\bar{q}^{\alpha}(x)\gamma_5 q^{\beta}(x) \leftrightarrow
\left.\dfrac{2}{z}\left(\dfrac{X-X^{\dag}}{2i}\right)^{\alpha\beta}(z,x)\right|_{z\rightarrow
0}$.

The pseudoscalar two-point correlator is the second variation of
the quadratic action with respect to the pseudoscalar sources.
(See the appendix {\bf \ref{pseudoscalar2ndvar}} for the details.)

\ba &&i\left\langle P_i(q)P_j(p)\right\rangle= \left.-i\
e^{-iS_{5D}[J_P]} \frac{\delta}{\delta
J^i_P(q)}\frac{\delta}{\delta J^j_P(p)}\
e^{iS_{5D}[J_P]}\right|_{J_P=0} =\frac{\delta}{\delta
J^i_P(q)}\frac{\delta}{\delta J^j_P(p)}\ S_{5D}[J_P]\nn
=\Lambda^2R^3\delta_{ij}\mathcal{D}(Q)\delta(p+q)+\ \mbox{terms
originating from the $X\leftrightarrow L^a_{\mu}, R^a_{\mu}$
mixing},\label{pp1}\ea

where $\mathcal{D}(Q)$ is the integral of the on-shell Lagrangian
with unity sources on the AdS boundary over $dz$, see
(\ref{Dapp}):

\be \mathcal{D}(Q) =
\frac{1}{4}\left[\frac{1}{\epsilon^2}+\frac{4\sigma}{m}+Q^2\log(Q^2\epsilon^2)\right].
\label{D}\ee

Here we have introduced the UV cutoff at $z=\epsilon$.

The parameter $\sigma$ is related to the quark condensate
(\ref{sigmaapp}):

\be C=\Lambda^2R^3\left(\frac{m}{4\epsilon^2}+\sigma\right).
\label{sigma}\ee

This allows us to express the pseudoscalar correlator in
(\ref{pp1}) through the condensate:

\ba i\left\langle
P_i(q)P_j(-q)\right\rangle=\delta_{ij}\left\{\frac{C}{m}+\frac{\Lambda^2R^3}{4}\
Q^2 \log (Q^2\epsilon^2) \right\}+\ \mbox{terms from the
$X\leftrightarrow L^a_{\mu}, R^a_{\mu}$ mixing}. \ea

The $X\leftrightarrow L^a_{\mu}, R^a_{\mu}$ mixing is proportional
to $Q^2$, so that when $Q^2=0$

\be i\left\langle
P_i(0)P_j(0)\right\rangle=\delta_{ij}\frac{C}{m}.\label{pseudoscalar2point}
\ee

One can see that the expression has the pole $\dfrac{1}{m}\propto
\dfrac{1}{m^2_{\pi}}$ due to the pion exchange.

Let us also consider a particular $N_f=2, m_u \neq m_d$ case. The
difference from the previous consideration will arise in the
expression of the bulk-to-boundary propagator and of the function
$\mathcal{D}(Q)=\dfrac{1}{4}\left[\dfrac{1}{\epsilon^2}+Q^2\log(Q^2\epsilon^2)\right]+\sigma\cdot
M^{-1}$ (c.f. (\ref{D})), where the matrices
$M=\mathrm{diag}(m_u,\ m_d), \sigma=\mathrm{diag}(\sigma_u,\
\sigma_d)\approx \sigma\cdot \mathbb{I}$. The second variation of
the action due to pseudoscalar sources gives:\ba i\left\langle
P_i(q)P_j(-q)\right\rangle&=&\delta_{ij}\frac{\Lambda^2R^3}{4}\
Q^2 \log (Q^2\epsilon^2)\delta(p+q) + 2\ \mathrm{tr}(t^i\Sigma
t^j)\ \delta(p+q)\nonumber\\ &+&\ \mbox{terms from the
$X\leftrightarrow L^a_{\mu}, R^a_{\mu}$ mixing}. \ea

The generalization of the result (\ref{pseudoscalar2point}) is the
following:

\be i\left\langle
P_i(0)P_j(0)\right\rangle=\delta_{ij}\frac{C}{m}-\frac{C\Delta
m}{2m^2}(\delta_{i0}\delta_{3j}+\delta_{j0}\delta_{3i}-i\delta_{i1}\delta_{2j}-i\delta_{j1}\delta_{2i}),\label{pseudoscalar2point03}
\ee where $\Delta m\equiv m_u-m_d$.

%%%%%%%%%%%%%%%%%%%%%%%%%%%%%%%%%%%%%%%%%%%%%%%%%%%%%%%%%%%%%%%%%%%%%%%%%%%%%%%%%%%%%%%%%%%%%%%%%%%%%%%%%
%%%%%%%%%%%%%%%%%%%%%%%%%%%%%%%%%%%%%%%%%%%%%%%%%%%%%%%%%%%%%%%%%%%%%%%%%%%%%%%%%%%%%%%%%%%%%%%%%%%%%%%%%
%%%%%%%%%%%%%%%%%%%%%%%%%%%%%%%%%%%%%%%%%%%%%%%%%%%%%%%%%%%%%%%%%%%%%%%%%%%%%%%%%%%%%%%%%%%%%%%%%%%%%%%%%
%%%%%%%%%%%%%%%%%%%%%%%%%%%%%%%%%%%%%%%%%%%%%%%%%%%%%%%%%%%%%%%%%%%%%%%%%%%%%%%%%%%%%%%%%%%%%%%%%%%%%%%%%

\subsection{Scalar currents\label{scalar}}

The model used in the previous subsection is insufficient when
dealing with the scalar currents due to their sensitivity to the
effective potential $V(X)$. Thus, in order to calculate the scalar
two-point correlator we will consider a soft-wall AdS/QCD model
\cite{softwall} in which $V(X)$ arises naturally. In this model
the dynamical fields have to vanish at $z\rightarrow\infty$ and
the solutions of the equations of motion are one-parametric unlike
(\ref{bulktoboundary}). This implies that in the absence of the
quartic term in the action the solution for the scalar field would
be proportional to its source -- the quark mass -- and would
vanish in the chiral limit. This would not allow us to disentangle
the spontaneous chiral symmetry breaking from the explicit one.
Introduction of the effective potential restores the correct
chiral limit and the constant $\kappa$ generates the mass
splitting between axial and vector mesons, $\kappa \approx 15$
being the best fit to the radial spectra of the axial mesons
\cite{softwall}. The parameter $\lambda$ in (\ref{dilaton}),
responsible for the slope of the Regge trajectory, is determined
to be $\lambda\approx 0.183\ \mathrm{GeV}^2$ \cite{softwall}.

We will use the action in the form \be S_{5D} = \int d^5x
\sqrt{g}e^{-\Phi}\Lambda^2 \textrm{tr} \left( |\partial
X|^2+\frac{3}{R^2}|X|^2
+\frac{\kappa}{R^2}|X|^4\right).\label{5dlagsoft}\ee

In the case when $X(z)$ is real and proportional to the unity
matrix, $X(z) = \dfrac{\mathcal{V}(z)}{2}\cdot \mathbb{I}$, the
nonlinear equation of motion assumes the form:

\be \dz\left(\frac{e^{-\Phi(z)} \dz\mathcal{V}(z)}{z^3}\right) +
\frac{e^{-\Phi(z)}}{z^5}\left(3\mathcal{V}(z) +
\frac{\kappa}{2}\mathcal{V}^3(z)\right)=0.\label{eomx}\ee

The AdS/QCD prescription (\ref{fields}) implies that
\be\mathcal{V}(z) \sim mz+\sigma z^3, z\rightarrow
0.\label{xsmallz}\ee

Eq. (\ref{eomx}) may be rewritten as an equation for the dilaton
\cite{softwall}:

\be \dz\Phi(z)=\frac{z^3}{\dz \mathcal{V}(z)}\left( \dz\left(
\frac{\dz \mathcal{V}(z)}{z^3} \right) +
\frac{3}{z^5}\mathcal{V}(z) + \frac{\kappa}{2z^5}\mathcal{V}^3(z)
\right),\label{eomdil}\ee from which it follows that
asymptotically \be\mathcal{V}(z) \sim \gamma z, \gamma =
2\sqrt{\dfrac{\lambda}{\kappa}}.\label{xlargez}\ee

The freedom in the behavior of the dilaton for small values of $z$
enables us to choose any form of the function $\mathcal{V}(z)$
provided that it satisfies the conditions
(\ref{xsmallz},\ref{xlargez}). For instance, \be \mathcal{V}(z) =
z\left(m+(\gamma-m)\tanh\left(\frac{\sigma
z^2}{\gamma-m}\right)\right)\label{xprofile}\ee is a suitable
choice. The dilaton will have an asymptotically parabolic profile
that switches from $\dfrac{1}{4}\kappa m^2 z^2, z\rightarrow 0$ to
$\lambda z^2, z\rightarrow \infty$ \cite{softwall}.

The nonlinear dependence of the classical solution
$\mathcal{V}(z)$ on the mass $m$ can be interpreted as a nonlinear
form of the bulk-to-boundary propagator in this model. More
exactly, the latter is be defined by the whole set of functions
$\mathcal{V}(z),\ \dfrac{\partial}{\partial m}\mathcal{V}(z),\
\dfrac{\partial^2}{\partial m^2} \mathcal{V}(z)$ etc. in the case
of a source that is uniform (i.e. its four-momentum equals zero)
and scalar (i.e. proportional to the identity matrix in the flavor
space $\mathbb{I}_{N_f\times N_f}$). Instead of a variation due to
a scalar source we will be differentiating the action with respect
to $m$.

The 5D on shell action (\ref{5dlagsoft}) equals:

\ba S_{5D\ Cl.} &=& \int
dz\frac{d^4q}{(2\pi)^4}N_f\Lambda^2R^3\left(
\frac{1}{4}\mathcal{V}\mathcal{O}_{5D}\mathcal{V}-\frac{1}{16}\tilde{\kappa}\mathcal{V}^4
\right) \ns &+& \int \frac{d^4q}{(2\pi)^4}N_f\Lambda^2R^3
\frac{1}{4}\mathcal{V}\mathcal{O}_{\partial\ 5D}\mathcal{V} \equiv
S_{vol} + S_{surf},\label{onshellaction1}\ea where
$\mathcal{O}_{5D} = \dz\left(\dfrac{e^{-\Phi(z)} \dz\
\cdot}{z^3}\right) + 3\dfrac{e^{-\Phi(z)}}{z^5}\ \cdot,\
\mathcal{O}_{\partial\ 5D} = \left.\dfrac{e^{-\Phi(z)} \dz}{z^3}\
\cdot\right|_{z=\epsilon\rightarrow 0},\ \tilde{\kappa} =
\kappa\dfrac{e^{-\Phi(z)}}{z^5}$.

On shell
$\mathcal{O}_{5D}\mathcal{V}-\dfrac{1}{2}\tilde{\kappa}\mathcal{V}^3=0$
(c.f. (\ref{eomx})) and \be S_{vol} =
\frac{1}{16}N_f\Lambda^2R^3\kappa\int dz
\frac{e^{-\Phi(z)}}{z^5}\mathcal{V}^4(z).\label{svol} \ee

The second derivative of the on-shell action (\ref{5dlagsoft})
with respect to the mass yields: \be i\left< S(0)S(0) \right> =
\frac{1}{2}\ \ddmass^2\mathcal{V}\ \frac{\delta S_{surf}}{\delta
X} + \frac{1}{4}\ \ddmass\mathcal{V}\ \frac{\delta^2
S_{surf}}{\delta X^2}\ \ddmass\mathcal{V} + \frac{1}{4}\
\ddmass\mathcal{V}\ \frac{\delta^2 S_{vol}}{\delta X^2}\
\ddmass\mathcal{V}.\label{sssoft}\ee

The details of the calculation can be found in the appendix {\bf
\ref{scalar2ndvar}}, eq. (\ref{ssterm1} -- \ref{ssterm3series}).

The result given in (\ref{ssapp}) is the following: \ba i\left<
S_i(0)S_j(0) \right> &=& \delta_{ij}\left\{
\frac{3}{8\pi^2}\lambda  N_c A_0 +
\frac{3}{16\pi^2}m\sqrt{\lambda\kappa} N_c A_1 +
\frac{N_c}{4\pi^2\epsilon^2}\right.\ns &+&\left.
\frac{3}{32\pi^2}m^2\kappa N_c \left( A_2-\frac{2}{3}+\log
\left(\frac{2\pi^2 C
\epsilon^2}{N_c}\sqrt{\frac{\kappa}{\lambda}}\right)-\frac{N_c}{2\pi^2
C}\sqrt{\frac{\lambda^3}{\kappa}} A_0 \right)\right\},\label{ss}
\ea where numerically \be A_0 = 0.377,\ A_1 = 0.977,\ A_2 =
-1.487.\ee

%%%%%%%%%%%%%%%%%%%%%%%%%%%%%%%%%%%%%%%%%%%%%%%%%%%%%%%%%%%%%%%%%%%%%%%%%%%%%%%%%%%%%%%%%%%%%%%%%%%%%%%%%
%%%%%%%%%%%%%%%%%%%%%%%%%%%%%%%%%%%%%%%%%%%%%%%%%%%%%%%%%%%%%%%%%%%%%%%%%%%%%%%%%%%%%%%%%%%%%%%%%%%%%%%%%
%%%%%%%%%%%%%%%%%%%%%%%%%%%%%%%%%%%%%%%%%%%%%%%%%%%%%%%%%%%%%%%%%%%%%%%%%%%%%%%%%%%%%%%%%%%%%%%%%%%%%%%%%
%%%%%%%%%%%%%%%%%%%%%%%%%%%%%%%%%%%%%%%%%%%%%%%%%%%%%%%%%%%%%%%%%%%%%%%%%%%%%%%%%%%%%%%%%%%%%%%%%%%%%%%%%

\section{The quartic pion Lagrangian\label{chirallag}}

The 5D effective action of the AdS/QCD model can be interpreted
not only as the generating functional for the correlators of the
QCD currents but also as a low-energy action of mesons. Indeed, if
the five-dimensional fields on the boundary are sources of the QCD
currents and these currents in their turn are sources of the
corresponding mesons, one can Kaluza--Klein decompose the fields
in the bulk and obtain an effective action for the modes of the 4D
boundary fields. These modes are proportional to the wavefunctions
of the mesons with corresponding quantum numbers. Integrating out
all the dynamics  along the $z$ axis we will obtain an effective
low-energy action for the mesons which may be rewritten as a sum
of the lowest-order terms of the Chiral Lagrangian. Since both
Lagrangians possess the same symmetries we shall assume that one
obtained from the AdS/QCD point of view reproduces the QCD Chiral
Lagrangian.

The NLO part of the Chiral Lagrangian that we are interested in is
the following (see, e.g., \cite{leutwyler}):

\ba \mathcal{L}_{\chi\ NLO} &=& L_1 \textrm{Tr}^2\left(\dmu
U^{\dag}\dmuu U\right) + L_2 \mathrm{Tr}\left(\dmu U^{\dag}\dnu
U\right)\mathrm{Tr}\left(\dmuu U^{\dag}\dnuu U\right) + L_3
\mathrm{Tr}\left( \dmu U^{\dag}\dmuu U\dnu U^{\dag}\dnuu U \right)
\ns &+& L_4
\mathrm{Tr}(D_{\mu}U^{\dag}D^{\mu}U)\mathrm{Tr}(U^{\dag}\chi+\chi^{\dag}U).
\label{chirallagexplicit}\ea

We will focus on the pion dynamics and will therefore need to
consider the axial fields in the bulk.

Let us consider the gauge sector of the hard-wall AdS/QCD action
(c.f. (\ref{full5daction})): \ba S_{5D}=\int d^5x
\sqrt{g}\frac{1}{g_5^2}\textrm{tr}\left(
-\frac{1}{4}(F_L^2+F_R^2)\right), \label{gaugeaction}\ea where
$F_L=dL-iL\wedge L,\ F_R=dR-iR\wedge R,\ DX=dX-iLX+iXR,
X(z)=\dfrac{1}{2}v(z)\cdot 1_{N_f\times N_f}, v(z)\equiv mz+\sigma
z^3$. We can redefine the action in terms of the vector and axial
fields $L=V+A, R=V-A$ and chose a gauge $V_z=A_z=\dmuu V_{\mu}=0$.
The $A_{\mu}$ field is divided into transverse and longitudinal
parts: $A_{\mu}=A_{\bot\mu}+\dmu\phi$.

Since for the axial and pseudoscalar currents $\dmu
\mathcal{A}^{\mu} = 2m P$, then $-2m\phi(z=0,x)$ is the source of
the pseudoscalar current. In its turn, the pseudoscalar current
being the source of pions, $\left< 0|P^a(p)|\pi^b(q) \right> =
iG_{\pi}\delta^{ab}\delta(p-q)=i\dfrac{C}{f_{\pi}}\
\delta^{ab}\delta(p-q)$, would suggest that $\phi^a(z=0,x)$ is
proportional to $\pi^a(x)$. We will establish the exact
proportionality later.

We can Kaluza-Klein decompose the fields $V$ and $\phi$: \be
V^a_{\mu}(z,x)=\sum_n V^{a(n)}_{\mu}(x) f^{(n)}_V(z),\
\phi^a(z,x)=\sum_n \phi^{a(n)}(x)
f^{(n)}_{\phi}(z).\label{kkmodes}\ee The normalization is the
following: \be \int \frac{dz}{z}
f^{(n)}(z)f^{(m)}(z)=\delta^{mn}.\label{norm} \ee

The functions $V^{a(n)}_{\mu}(x)$ and $\phi^{a(n)}(x)$ are
proportional to the four-dimensional wavefunctions of the vector
and pseudoscalar mesons respectively. We shall focus on the lowest
mode $\phi^{a(0)}(x)$ which corresponds to the pion field
$\pi^a(x)$.

The functions $f(z)$ in their turn are the solutions of the
equations of motion with the boundary conditions $f(0)=\dz
f(z_m)=0$. The lowest modes $f_{\phi}(z),\ f_V(z)$ correspond to
the $\pi$ and $\rho$ mesons. $f_V(z)$ has the following form: \ba
&&f_V(z)=N_V zI_1(m_{\rho}z),\ N_V^{-2}=\frac{z_m^2}{2}\left(
I_1^2(m_{\rho}z_m) - I_2(m_{\rho}z_m)I_0(m_{\rho}z_m)
\right).\label{vmode}\ea

After we integrate out all the dynamics along the $z$ axis and
leave only the lowest mode of the field $\phi(x)$ the
five-dimensional action (\ref{gaugeaction}) yields:

\ba S_{5D} &\rightarrow& \int d^4x\ \textrm{tr}\left(
N^2_{\pi}\cdot \frac{1}{2}\dmu \phi^{(0)}(x) \dmuu\phi^{(0)}(x)+
g_{\phi^4}\cdot[\dmu\phi^{(0)}(x),\dnu\phi^{(0)}(x)][\dmuu\phi^{(0)}(x),\dnuu\phi^{(0)}(x)]
\right.\ns &+&\left.\phantom{\frac{1}{2}} g_{m\ \phi^2}\cdot m\dmu
\phi^{(0)}(x) \dmuu\phi^{(0)}(x) \right). \label{chiralfromads}\ea

The quantities $N_{\pi},\ g_{\phi^4},\ g_{m\ \phi^2}$ are the
corresponding integrals over $z$ and their explicit expressions
can be found in the appendices {\bf \ref{chiralparnorm}} eq.
(\ref{Npi}), {\bf \ref{chiralpar}} eq. (\ref{gphi4}) and {\bf
\ref{chiralparnorm}} eq. (\ref{gmphi2}) respectively.

The canonical normalization of the pion field in
(\ref{chiralfromads}) demands that the proportionality between the
pion and the $\phi$ assume the form: \be
N_{\pi}\phi^{a(0)}(x)=F_{\pi}\pi^a(x).\label{propoptionality}\ee

There are four terms in the next-to-leading order Chiral
Lagrangian (\ref{chirallagexplicit}) that are relevant to us: \ba
&& \mathcal{P}_1 = \textrm{Tr}^2\left(\dmu U^{\dag}\dmuu U\right),
\nn \mathcal{P}_2 = \mathrm{Tr}\left(\dmu U^{\dag}\dnu
U\right)\mathrm{Tr}\left(\dmuu U^{\dag}\dnuu U\right), \nn
\mathcal{P}_3 = \mathrm{Tr}\left( \dmu U^{\dag}\dmuu U\dnu
U^{\dag}\dnuu U \right),\nn \mathcal{P}_4 =
\mathrm{Tr}(D_{\mu}U^{\dag}D^{\mu}U)\mathrm{Tr}(U^{\dag}\chi+\chi^{\dag}U)=
8\frac{\Sigma}{F^2_{\pi}}m \dmu \pi^a(x)\dmuu \pi^a(x) +
\mathcal{O}(\pi^4).\label{chiralterms}\ea

Analogously to \cite{sakaisugimoto2} the Skyrme--like quartic term
in (\ref{chiralfromads}) induced by the AdS dynamics is the lowest
power in $\pi$ of a linear combination of $\mathcal{P}_i$: \ba
&&g_{\phi^4}\frac{F^4_{\pi}}{N^4_{\pi}}[\dmu\pi(x),\dnu\pi(x)][\dmuu\pi(x),\dnuu\pi(x)]
= \sum_{i=1}^3 L_i \mathcal{P}_i + o(\pi^4, p^4),\nn \mbox{where}\
L_1=-\frac{g_{\phi^4}}{8}\frac{F^4_{\pi}}{N^4_{\pi}},\
L_2=-\frac{g_{\phi^4}}{4}\frac{F^4_{\pi}}{N^4_{\pi}}=2L_1,\
L_3=\frac{3g_{\phi^4}}{4}\frac{F^4_{\pi}}{N^4_{\pi}}=-6L_1.\label{l123}
\ea The relation between $L_1, L_2, L_3$ is in agreement with
\cite{kim,sakaisugimoto2}, but, unlike in \cite{kim}, we did not
need to consider the $\rho\pi\pi$ interaction.

From (\ref{chiralfromads}) and (\ref{chiralterms}) we obtain (see
(\ref{appl4})): \be L_4=\frac{F^4_{\pi}}{8\Sigma}g_{m\ \phi^2}
N^{-2}_{\pi}.\label{l4}\ee

One can explicitly demonstrate that $L_i$ are regular in the
chiral limit and are $\propto N_c$.

%%%%%%%%%%%%%%%%%%%%%%%%%%%%%%%%%%%%%%%%%%%%%%%%%%%%%%%%%%%%%%%%%%%%%%%%%%%%%%%%%%%%%%%%%%%%%%%%%%%%%%%%%
%%%%%%%%%%%%%%%%%%%%%%%%%%%%%%%%%%%%%%%%%%%%%%%%%%%%%%%%%%%%%%%%%%%%%%%%%%%%%%%%%%%%%%%%%%%%%%%%%%%%%%%%%
%%%%%%%%%%%%%%%%%%%%%%%%%%%%%%%%%%%%%%%%%%%%%%%%%%%%%%%%%%%%%%%%%%%%%%%%%%%%%%%%%%%%%%%%%%%%%%%%%%%%%%%%%
%%%%%%%%%%%%%%%%%%%%%%%%%%%%%%%%%%%%%%%%%%%%%%%%%%%%%%%%%%%%%%%%%%%%%%%%%%%%%%%%%%%%%%%%%%%%%%%%%%%%%%%%%

\section{Spectral density of the Dirac operator\label{spectrum}}

While both the correlation functions and the Chiral Lagrangian
parameters may be obtained from the effective 5D action more or
less straightforwardly, the spectral density of the Dirac operator
is a more complex entity because of its fermionic nature. In order
to be able to calculate it in dual holographic theories we have to
express it through the partition function.

Let us consider a Yang-Mills theory of gluons $A_{\mu}$ with $N_c$
colors and $N_f$ flavors of quarks $\Psi$ in the fundamental
representation of $SU(N_c)$ with mass $m$ and coupling $g$. We
define the Dirac operator $\hat{D}$ as follows: \be \hat{D} \equiv
\gm(\dmu+igA_{\mu}). \ee

Let $\{\lambda_n\}$ be the eigenvalues of the Dirac operator: \be
\mbox{for}\ \forall\lambda \in \{\lambda_n\}\ \exists \Psi:\
i\hat{D}\Psi = \lambda\Psi.\ee The spectrum is $N_f$ times
degenerate due to the global $SU(N_f)_V$ symmetry. $\hat{D}$ and
$\gamma_5$ anticommute, so that for every eigenvalue $\lambda_n$
and eigenvector $\Psi_n$ there is eigenvalue $-\lambda_n$
corresponding to $\gamma_5 \Psi_n$.

We can define the Euclidean four-volume $V$ and the corresponding
spectral density of the Dirac operator: \be \rho(\lambda) \equiv
\frac{1}{V}\left\langle
\sum_n\delta(\lambda-\lambda_n)\right\rangle_A.\ee

We shall introduce a regulatory mass $\mu$ and rewrite an infinite
sum in terms of functional trace. The density was originally
expressed in terms of a resolvent in \cite{efetov}.

\ba &&\rho(\lambda) = \frac{1}{V}\left\langle
\sum_n\delta(\lambda-\lambda_n)\right\rangle_A = \frac{1}{\pi
V}\left\langle \lim_{\mu\rightarrow 0} \sum_n \frac{\mu}{\mu^2 +
(\lambda-\lambda_n)^2} \right\rangle_A \nn= \frac{-i}{2\pi V}
\lim_{\mu\rightarrow 0}\left\langle Tr[i\hat{D}-\lambda-i
\mu]^{-1} - Tr[i\hat{D}-\lambda+i \mu]^{-1}\right\rangle_A \nn=
\frac{1}{2\pi V}\lim_{\mu\rightarrow 0}\dm \left\langle \log
Det[i\hat{D}-\lambda-i \mu] + \log Det [i\hat{D}-\lambda+i \mu]
\right\rangle_A \ea

Using a procedure called the "replica trick" ($\log z = \left.\dn
z^n\right|_{n=0}$) we will be able to rewrite the logarithm of a
determinant in terms of a partition function with the introduction
of ghost quarks $\chi$. Similar calculations can be seen in
\cite{replica}.

\ba &&\rho(\lambda) = \frac{1}{\pi V}\lim_{\mu\rightarrow 0}\dm\
\Re \left\langle \log \int \prod_{I=1}^{N_f} D\chi_I D\bar{\chi}_I
\exp\left\{-\int d^4x\
\bar{\chi}_I(i\hat{D}-\lambda-i\mu)\chi_I\right\}\right\rangle_A(N_f\
\chi\ \textrm{flavors}) \nn= \frac{1}{\pi V}\lim_{\mu\rightarrow
0}\dm \lim_{n\rightarrow 0}\dn\ \Re \left\langle \int
\prod_{I=1}^{n\cdot N_f} D\chi_I D\bar{\chi}_I \exp\left\{-\int
d^4x\
\bar{\chi}_I(i\hat{D}-\lambda-i\mu)\chi_I\right\}\right\rangle_A\nn=
\frac{1}{\mathcal{Z}_{QCD}\pi V}\lim_{\mu\rightarrow 0}\dm
\lim_{n\rightarrow 0}\dn\ \Re \int DA \prod_{J=1}^{N_f}D\Psi_J
D\bar{\Psi}_J \prod_{I=1}^{n\cdot N_f} D\chi_I
D\bar{\chi}_I\nn\times \exp\left\{i\int d^4x\
\bar{\chi}_I(i\hat{D}-\lambda-i\mu)\chi_I +
\bar{\Psi}_J(i\hat{D}-m)\Psi_J -\frac{1}{2}\textrm{tr}\
F_{\mu\nu}F^{\mu\nu} + \frac{g^2\theta}{32\pi^2}\textrm{tr}\
F_{\mu\nu}\tilde{F}^{\mu\nu}\right\}.\label{spectraldensity} \ea

For further simplicity we will introduce the following effective
actions: \ba &&S_{QCD} \equiv -i\log \mathcal{Z}_{QCD}= -i\log\int
DA \prod_{J=1}^{N_f}D\Psi_J D\bar{\Psi}_J \nn
\times\exp\left\{i\int d^4x\ \bar{\Psi}_J(i\hat{D}-m)\Psi_J
-\frac{1}{2}\textrm{tr}\ F_{\mu\nu}F^{\mu\nu} +
\frac{g^2\theta}{32\pi^2}\textrm{tr}\
F_{\mu\nu}\tilde{F}^{\mu\nu}\right\},\label{sqcd}\\
&&S_{QCD+ghosts}\equiv -i\log \mathcal{Z}_{QCD+ghosts}= -i\log\int
DA \prod_{J=1}^{N_f}D\Psi_J D\bar{\Psi}_J \prod_{I=1}^{n\cdot N_f}
D\chi_I D\bar{\chi}_I\ns &&\times \exp\left\{i\int d^4x\
\bar{\chi}_I(i\hat{D}-\mu-i\lambda)\chi_I +
\bar{\Psi}_J(i\hat{D}-m)\Psi_J -\frac{1}{2}\textrm{tr}\
F_{\mu\nu}F^{\mu\nu} + \frac{g^2\theta}{32\pi^2}\textrm{tr}\
F_{\mu\nu}\tilde{F}^{\mu\nu}\right\}.\label{sqcdgh} \ea

The expression (\ref{spectraldensity}) may now be simplified: \ba
\rho(\lambda,m) &=& \left.\frac{1}{\pi V}\ \Re\
e^{-iS_{QCD}}\left( -\dm S \dn S +\ i \dm\dn S
\right)e^{iS_{QCD+ghosts}} \right|_{\mu=n=0}\ns &=&
\left.\frac{1}{\pi V}\ \Re\ \left( -\dm \cdot\dn \cdot\ +\ i\
\dm\dn\
\cdot\right)S_{QCD+ghosts}(\lambda,\mu,n,m)\right|_{\mu=n=0}.\label{recipe}\ea

%%%%%%%%%%%%%%%%%%%%%%%%%%%%%%%%%%%%%%%%%%%%%%%%%%%%%%%%%%%%%%%%%%%%%%%%%%%%%%%%%%%%%%%%%%%%%%%%%%%%%%%%%
%%%%%%%%%%%%%%%%%%%%%%%%%%%%%%%%%%%%%%%%%%%%%%%%%%%%%%%%%%%%%%%%%%%%%%%%%%%%%%%%%%%%%%%%%%%%%%%%%%%%%%%%%
%%%%%%%%%%%%%%%%%%%%%%%%%%%%%%%%%%%%%%%%%%%%%%%%%%%%%%%%%%%%%%%%%%%%%%%%%%%%%%%%%%%%%%%%%%%%%%%%%%%%%%%%%
%%%%%%%%%%%%%%%%%%%%%%%%%%%%%%%%%%%%%%%%%%%%%%%%%%%%%%%%%%%%%%%%%%%%%%%%%%%%%%%%%%%%%%%%%%%%%%%%%%%%%%%%%

\subsection{Spectral density in hard wall AdS/QCD\label{spectrumads}}

Let us denote the physical quarks $\Psi_J$, $J=1...N_f$, and the
ghost quarks $\chi_I$, $I=1...n\cdot N_f$ (See formula
\ref{spectraldensity}). We use the AdS/QCD prescription
(\ref{fields}) to put a five-dimensional scalar field
$X^{\alpha\beta}(z,x)$, where $\alpha, \beta = 1...(n+1)N_f$, into
correspondence with the scalar currents: \ba
&&\left.\dfrac{2}{z}X^{\alpha\beta}(z,x)\right|_{z\rightarrow 0}
\leftrightarrow \bar{\Psi}_L^{\alpha}(x) \Psi_R^{\beta}(x),\
\alpha,\beta = 1...N_f,\label{psipsi}\\
&&\left.\dfrac{2}{z}X^{\alpha(\beta+N_f)}(z,x)\right|_{z\rightarrow
0} \leftrightarrow \bar{\Psi}_L^{\alpha}(x) \chi_R^{\beta}(x),\
\alpha = 1...N_f,\ \beta=1...nN_f,\label{psichi}\\
&&\left.\dfrac{2}{z}X^{(N_f+\alpha)\beta}(z,x)\right|_{z\rightarrow
0} \leftrightarrow \bar{\chi}_L^{\alpha}(x) \Psi_R^{\beta}(x),\
\alpha = 1...nN_f,\ \beta=1...N_f,\label{chipsi}\\
&&\left.\dfrac{2}{z}X^{(N_f+\alpha)(N_f+\beta)}(z,x)\right|_{z\rightarrow
0} \leftrightarrow \bar{\chi}_L^{\alpha}(x) \chi_R^{\beta}(x),\
\alpha = 1...nN_f,\ \beta=1...nN_f.\label{chichi} \ea

The Lagrangian (\ref{spectraldensity}) suggests that the
Lagrangian of the five-dimensional model possesses a
$SU(N_f)\times SU(nN_f)$ gauge symmetry. The action of the scalar
field $X$ is (\ref{full5daction}): \be S_{5D} = \int d^5x
\sqrt{g}\Lambda^2 \textrm{tr} \left( |\partial
X|^2+\frac{3}{R^2}|X|^2 \right).\label{5dl}\ee

The source for $X$ is: $$
\frac{2}{\epsilon}X^{\alpha\beta}(\epsilon,x) =
\mathrm{diag}(\underbrace{m\ ...\ m}_{N_f\ \mbox{раз}},\
\underbrace{\lambda+i\mu\ ...\ \lambda+i\mu}_{nN_f\ \mbox{раз}})
\equiv m\cdot \mathbb{P}^{\Psi} +(\lambda+i\mu)\cdot
\mathbb{P}^{\chi} ,\label{sourcexgh}
$$ where $\mathbb{P}^{\Psi} \equiv \delta_{\alpha\beta}\ \mbox{iff}\
\alpha,\beta=1...N_f$ is a projector on the quark states
$\Psi^{\alpha}$ in flavor space, $\mathbb{P}^{\chi} \equiv
\delta_{\alpha\beta}\ \mbox{iff}\ \alpha,\beta=N_f+1...(n+1)N_f$
is an analogous projector on the ghost states $\chi^{\beta}$.

The function $X(z,x)$ in the bulk is expressed in terms of the
source via the bulk-to-boundary propagator, similar to
(\ref{bulktoboundary}):

\ba X_{\alpha\beta}(z,x)&=&\int \frac{d^4q}{(2\pi)^4}\ e^{-iqx}
\mathcal{K}_{\alpha\gamma}(z,q)\frac{2}{\epsilon}X_{\gamma\beta}(\epsilon,q)\nonumber\\
&=&\int \frac{d^4q}{(2\pi)^4} e^{-iqx}
\mathcal{K}_{\alpha\gamma}(z,q)
\delta(q)\left[m\mathbb{P}^{\Psi}_{\gamma\beta} +
(\lambda+i\mu)\mathbb{P}^{\chi}_{\gamma\beta}\right]\label{xbulkgh}\ea
(c.f. (\ref{xbulk})).

We assume that the bulk-to-boundary has the following form: \be
\mathcal{K}_{\alpha\gamma}(z,q)=\mathcal{K}(z,q)\mathbb{P}^{\Psi}_{\alpha\gamma}+\mathcal{\Tilde{K}}(z,q)\mathbb{P}^{\chi}_{\alpha\gamma}.
\label{bulktoboundarygh} \ee

In QCD with quarks $\Psi$ with mass $m$ and condensate
$\Sigma=CN_f=\sigma N_f\Lambda^2R^3$ the field $X$ in the bulk
equals $X(z)=\dfrac{1}{2}\left(mz+\sigma
z^3\right)\equiv\dfrac{1}{2}v(z)$. A QCD-like theory with ghosts
$\chi$ with mass $\lambda+i\mu$ should have the field $X$ of the
same form: $X(z)=\dfrac{1}{2}\left((\lambda+i\mu)z+\Tilde{\sigma}
z^3\right)$, where
$\Tilde{\sigma}=\dfrac{\Tilde{C}}{\Lambda^2R^3}$, $\Tilde{C}$
being the condensate of ghosts. In the first approximation
$\Tilde{C}=C$, but this equality is imprecise because of the
mass-dependence of the condensate. This leads us to a conclusion
that the quantity $\mathcal{\Tilde{D}}(Q)$ (analogous to
$\mathcal{D}(Q)$ from subsection {\bf \ref{2point}} eq. (\ref{D})
for quarks $\Psi$ where that quantity was defined as the integral
over $dz$ of the on-shell Lagrangian with unity sources on the AdS
boundary) equals: \be
\mathcal{\Tilde{D}}(Q)=\frac{\Tilde{C}}{\Lambda^2R^3(\lambda+i\mu)}+\frac{Q^2}{4}\
\log{(Q^2\epsilon^2)}.\label{Dgh}\ee

From Lagrangian in (\ref{spectraldensity}) it follows that the
current $\bar{\chi}\chi$ has to have a complex-valued source,
while $\bar{\chi}\gamma_5\chi$ has to have none. This implies that
$X(\epsilon,x)$ has to be complex and
$X(\epsilon,x)-X^{\dag}(\epsilon,x)=0$. The only way to do this is
to temporarily introduce another field $Y$ instead of $X^{\dag}$
and to treat $X$ and $Y$ independently, so that in general
$Y^{\dag}\neq X$, in fact, $Y=X$.

%to treat $X$ and $X^{\dag}$ fields independently, i.e., $X^{\dag}
%\neq \bigl(X\bigr)^{\dag}$. In fact, $X^{\dag}$ has to be equal to
%$X$.

The action may be expressed in terms of the sources: \ba
S_{5D}&&=\Lambda^2R^3 \mathrm{tr} \int \frac{d^4q}{(2\pi)^4}
\left[\mathcal{D}(Q)\mathbb{P}^{\Psi} +
\mathcal{\Tilde{D}}(Q)\mathbb{P}^{\chi}\right]\nn
\times\left[m\mathbb{P}^{\Psi} +
(\lambda+i\mu)\mathbb{P}^{\chi}\right]\delta(q)
\times\left[m\mathbb{P}^{\Psi} +
(\lambda+i\mu)\mathbb{P}^{\chi}\right]\delta(q)\nn
=\Lambda^2R^3\delta(0)\left(m^2N_f\mathcal{D}(0)+(\lambda+i\mu)^2nN_f\mathcal{\Tilde{D}}(0)\right)
=N_fV\left(m C+(\lambda+i\mu)n\Tilde{C}\right).\label{actiongh}\ea

To calculate the spectral density we will use the formula
(\ref{recipe}). Since $S_{5D}$ is a linear function of $\mu\cdot
n$, the first term $\left.\dm S \dn S\right|_{n=0}=0$. Hence, \ba
\rho(\lambda) = \left.\frac{1}{\pi V}\Re\ i\dm\dn
S_{5D}\right|_{\mu=n=0} = \left.\frac{1}{\pi V}\ \Re\ N_f V
\left[-\Tilde{C} + i(\lambda+i\mu)\dm\Tilde{C}
\right]\right|_{\mu=0}.\label{spectraldensityads1}\ea

Let us denote $C(m)=\sum_{n=0}^{\infty}C_n m^n$ for $\Re(m)>0$.
The expression (\ref{spectraldensityads1}) yields:\be
\rho(\lambda) = -\frac{N_f}{\pi} \sum_{n=0}^{\infty}
C_{n}(n+1)\lambda^{n} =-\frac{1}{\pi}\Sigma_0 -\frac{1}{\pi}
\sum_{n=1}^{\infty} \Sigma_{n}(n+1)\lambda^{n}\
(\Re(\lambda)>0),\label{spectraldensityads2}\ee where
$\Sigma(m)=\sum\limits_{n=0}^{\infty}\Sigma_n m^n\ \mbox{for}\
\Re(m)>0,\ \Sigma_0 \equiv \Sigma|_{m=0}$.

Making use of the formula from \cite{sumrules} that describes the
mass-dependence of the quark condensate $\Sigma(m)$ \be \Sigma(m)
= \Sigma_0\left(1-\frac{3m^2_{\pi}\log
m^2_{\pi}/\mu^2_{hadr}}{32\pi^2F^2_{\pi}}+...\right),\ee we
obtain: \be \rho(\lambda) = -\frac{1}{\pi}\Sigma_0\left( 1 -
\frac{3\Sigma_0}{8N_f\pi^2F^4_{\pi}}|\lambda|
-\frac{3\Sigma_0}{4N_f\pi^2F^4_{\pi}}|\lambda|\log|\lambda/\tilde{\mu}_{hadr}|\right).
\label{spectraldensityadsfinal}\ee

As one may notice, firstly, the result satisfies the Casher--Banks
identity \cite{casherbanks}. Secondly, it reproduces up to the
factor $\propto N_f^2-4$ a well-known formula from QCD, first
derived in \cite{smilgastern}, \be\frac{\rho'(0)}{\rho(0)}\propto
\frac{\Sigma}{F_{\pi}^4}\label{densitylargeN}\ee for the term
linear in $\lambda$.

Nevertheless, eq. (\ref{spectraldensityadsfinal}) does not
describe the dependence of the spectral density on the mass $m$
and does not include the terms $\lambda^2$ and higher powers of
$\lambda$.

%%%%%%%%%%%%%%%%%%%%%%%%%%%%%%%%%%%%%%%%%%%%%%%%%%%%%%%%%%%%%%%%%%%%%%%%%%%%%%%%%%%%%%%%%%%%%%%%%%%%%%%%%
%%%%%%%%%%%%%%%%%%%%%%%%%%%%%%%%%%%%%%%%%%%%%%%%%%%%%%%%%%%%%%%%%%%%%%%%%%%%%%%%%%%%%%%%%%%%%%%%%%%%%%%%%
%%%%%%%%%%%%%%%%%%%%%%%%%%%%%%%%%%%%%%%%%%%%%%%%%%%%%%%%%%%%%%%%%%%%%%%%%%%%%%%%%%%%%%%%%%%%%%%%%%%%%%%%%
%%%%%%%%%%%%%%%%%%%%%%%%%%%%%%%%%%%%%%%%%%%%%%%%%%%%%%%%%%%%%%%%%%%%%%%%%%%%%%%%%%%%%%%%%%%%%%%%%%%%%%%%%

\section{Low-energy theorems from the holographic point of view\label{compatibility}}

Having obtained in sections {\bf \ref{2point} -- \ref{spectrum}}
the expressions for all the necessary ingredients of the
low-energy theorems in the AdS/QCD framework we shall determine
now whether (and how accurately) these theorems are satisfied in
the dual models.

We will be using the following result for the two-point correlator
of the topological charge density \cite{wveneziano}:

\be \frac{\int d^4x d^4y \left< Q(x)Q(y) \right> }{V} =
\frac{mBF^2_{\pi}}{2}+o(m)=mC +o(m),\label{topol2point}\ee that
was reproduced in AdS/QCD in \cite{u1primer}.

Let us consider the first low-energy theorem (\ref{sspp}). One can
see that all its three sides have a pole at $m\rightarrow 0$, and
the pole residue in the left-hand side $C$
(\ref{pseudoscalar2point}) equals that of the central side.
Indeed,

\be
\dfrac{G^2_{\pi}}{m^2_{\pi}}=\dfrac{4C^2}{m^2_{\pi}F^2_{\pi}}=\dfrac{C}{m}.\label{therorem1lm}\ee
As for the right-hand side, from (\ref{spectraldensityads2}) we
obtain \be \int d\lambda \frac{4m^2
\rho(\lambda,m)}{(\lambda^2+m^2)^2} = \frac{4Cm^2}{\pi}\int
d\lambda \frac{1}{(\lambda^2+m^2)^2} +
\mathcal{O}(1)=\frac{C}{m}+\mathcal{O}(1),\ m\rightarrow
0,\label{therorem1r}\ee which agrees with the residues of the
left-hand and central sides. As for the finite part of the
left-hand and central sides, one can straightforwardly check that
both $\left< S_i(0)S_j(0) \right>$ (\ref{ss}) and
$B^2(L_3-4L_4+3)$ (\ref{l123}, \ref{l4}) are $\propto N_c$.

In the case of the second low-energy theorem (\ref{ppss}) one can
also observe the similar analytical structure of the left- and
right-hand sides at the $m\rightarrow 0$ pole, the residues are
again in agreement (see (\ref{therorem1lm}, \ref{topol2point}))

As for the third low-energy theorem (\ref{pp}), one can see that
the left-hand side defined by (\ref{pseudoscalar2point03}) и
(\ref{topol2point}), coincides with the right-hand side when
$m\rightarrow 0$: \be \frac{2Cm\Delta m}{\pi}\int d\lambda
\frac{1}{(\lambda^2+m^2)^2} - \frac{\Delta m}{m^3}Cm +
\mathcal{O}(1)=-\frac{C\Delta
m}{2m^2}+\mathcal{O}(1).\label{therorem3}\ee

%%%%%%%%%%%%%%%%%%%%%%%%%%%%%%%%%%%%%%%%%%%%%%%%%%%%%%%%%%%%%%%%%%%%%%%%%%%%%%%%%%%%%%%%%%%%%%%%%%%%%%%%%
%%%%%%%%%%%%%%%%%%%%%%%%%%%%%%%%%%%%%%%%%%%%%%%%%%%%%%%%%%%%%%%%%%%%%%%%%%%%%%%%%%%%%%%%%%%%%%%%%%%%%%%%%
%%%%%%%%%%%%%%%%%%%%%%%%%%%%%%%%%%%%%%%%%%%%%%%%%%%%%%%%%%%%%%%%%%%%%%%%%%%%%%%%%%%%%%%%%%%%%%%%%%%%%%%%%
%%%%%%%%%%%%%%%%%%%%%%%%%%%%%%%%%%%%%%%%%%%%%%%%%%%%%%%%%%%%%%%%%%%%%%%%%%%%%%%%%%%%%%%%%%%%%%%%%%%%%%%%%

\section{Discussion}

As we can see, simple hard-wall or soft-wall AdS/QCD models allow
us to test their compatibility with the low-energy-theorems of
QCD. Indeed, using the dual description of the quantum field
theory in the strong coupling regime we have demonstrated that the
theorems are satisfied in the chiral limit when the quark and pion
masses tend to zero. The zero-momentum two-point correlators of
the pseudoscalar currents precisely coincide with the corollary of
the theorems (\ref{therorem1lm}, \ref{therorem1r},
\ref{therorem3}), and moreover the result
(\ref{pseudoscalar2point}) fort the correlation function
$\left<P_i(0)P_j(0)\right>$ independently agrees with the field
theory result (\ref{therorem1lm}).

Holographic models also induce low-energy dynamics of
four-dimensional fields on the AdS boundary that can be put into
correspondence with meson wavefunctions with suitable quantum
numbers. This allows us to calculate the parameters of the
effective low-energy Chiral Lagrangian in those models
(\ref{l123}, \ref{l4}). The established relation between the
coefficients $L_1, L_2$ и $L_3$ (\ref{l123}) reproduces the result
\cite{sakaisugimoto2} in the "top-down" Sakai--Sugimoto model
\cite{sakaisugimoto1}.

Finally, we succeeded in expressing the spectral density of the
Dirac operator in terms of a partition function of a theory which
includes QCD and ghost quark fields (\ref{spectraldensity},
\ref{recipe}). This allowed us to calculate the density in the
AdS/QCD framework, the result (\ref{spectraldensityadsfinal})
agreeing with the Casher--Banks identity \cite{casherbanks} and
reproducing the general properties of the QCD formula
\cite{smilgastern} in the linear order in $\lambda$.

Being for the most part simplified, AdS/QCD models such as
\cite{hardwall}, \cite{softwall} and \cite{softwall1} need further
improvement and sharpening. One of their major shortcomings is the
significant dependence of the results on the particular choice of
the background geometry, presence of the dilaton and its profile,
form of the effective potentials etc. For instance, the two-point
correlator of scalar currents calculated in the hard-wall model
\cite{hardwall}, significantly differs from the expression
(\ref{ss}) and has an unphysical pole structure. On the other
hand, the soft-wall model \cite{softwall} has not been yet
directly generalized to describe quark flavors with different
masses, and therefore in its present form does not allow us to
calculate the spectral density using the formula (\ref{recipe}).
We intend to study the dependence of the AdS/QCD results for the
coefficients of the NLO Chiral Lagrangian on the choice of the
model (hard-wall/ soft-wall) elsewhere, although at the moment it
seems that the hard-wall model does not reproduce the coefficient
$L_7$.

All the aforementioned suggests that studying the AdS/QCD models
by means of testing them with exact equations established on the
field theory side of the correspondence is quite a promising task
and will hopefully lead to the refinement of those models.
Alongside with the advancement of the "top-down" approach that
produces more rigorous results at a cost of significant
complexity, this should reveal us the possible structure of the
ultimate dual of the QCD.

%%%%%%%%%%%%%%%%%%%%%%%%%%%%%%%%%%%%%%%%%%%%%%%%%%%%%%%%%%%%%%%%%%%%%%%%%%%%%%%%%%%%%%%%%%%%%%%%%%%%%%%%%
%%%%%%%%%%%%%%%%%%%%%%%%%%%%%%%%%%%%%%%%%%%%%%%%%%%%%%%%%%%%%%%%%%%%%%%%%%%%%%%%%%%%%%%%%%%%%%%%%%%%%%%%%
%%%%%%%%%%%%%%%%%%%%%%%%%%%%%%%%%%%%%%%%%%%%%%%%%%%%%%%%%%%%%%%%%%%%%%%%%%%%%%%%%%%%%%%%%%%%%%%%%%%%%%%%%
%%%%%%%%%%%%%%%%%%%%%%%%%%%%%%%%%%%%%%%%%%%%%%%%%%%%%%%%%%%%%%%%%%%%%%%%%%%%%%%%%%%%%%%%%%%%%%%%%%%%%%%%%

\section{Acknowledgments.}

I am thankful to A. S. Gorsky for formulating a very interesting
and rewarding problem, as well as for valuable discussions and
general guidance. This work was supported by the grants
RFBR-07-02-00878, NSh-3035.2008.2, MK-544.2009.1  and the Dmitry
Zimin Dynasty Foundation.

%%%%%%%%%%%%%%%%%%%%%%%%%%%%%%%%%%%%%%%%%%%%%%%%%%%%%%%%%%%%%%%%%%%%%%%%%%%%%%%%%%%%%%%%%%%%%%%%%%%%%%%%%
%%%%%%%%%%%%%%%%%%%%%%%%%%%%%%%%%%%%%%%%%%%%%%%%%%%%%%%%%%%%%%%%%%%%%%%%%%%%%%%%%%%%%%%%%%%%%%%%%%%%%%%%%
%%%%%%%%%%%%%%%%%%%%%%%%%%%%%%%%%%%%%%%%%%%%%%%%%%%%%%%%%%%%%%%%%%%%%%%%%%%%%%%%%%%%%%%%%%%%%%%%%%%%%%%%%
%%%%%%%%%%%%%%%%%%%%%%%%%%%%%%%%%%%%%%%%%%%%%%%%%%%%%%%%%%%%%%%%%%%%%%%%%%%%%%%%%%%%%%%%%%%%%%%%%%%%%%%%%

%%%%%%%%%%%%%%%%%%%%%%%%%%%%%%%%%%%%%%%%%%%%%%%%%%%%%%%%%%%%%%%%%%%%%%%%%%%%%%%%%%%%%%%%%%%%%%%%%%%%%%%%%
%%%%%%%%%%%%%%%%%%%%%%%%%%%%%%%%%%%%%%%%%%%%%%%%%%%%%%%%%%%%%%%%%%%%%%%%%%%%%%%%%%%%%%%%%%%%%%%%%%%%%%%%%
%%%%%%%%%%%%%%%%%%%%%%%%%%%%%%%%%%%%%%%%%%%%%%%%%%%%%%%%%%%%%%%%%%%%%%%%%%%%%%%%%%%%%%%%%%%%%%%%%%%%%%%%%
%%%%%%%%%%%%%%%%%%%%%%%%%%%%%%%%%%%%%%%%%%%%%%%%%%%%%%%%%%%%%%%%%%%%%%%%%%%%%%%%%%%%%%%%%%%%%%%%%%%%%%%%%

%%%%%%%%%%%%%%%%%%%%%%%%%%%%%%%%%%%%%%%%%%%%%%%%%%%%%%%%%%%%%%%%%%%%%%%%%%%%%%%%%%%%%%%%%%%%%%%%%%%%%%%%%
%%%%%%%%%%%%%%%%%%%%%%%%%%%%%%%%%%%%%%%%%%%%%%%%%%%%%%%%%%%%%%%%%%%%%%%%%%%%%%%%%%%%%%%%%%%%%%%%%%%%%%%%%
%%%%%%%%%%%%%%%%%%%%%%%%%%%%%%%%%%%%%%%%%%%%%%%%%%%%%%%%%%%%%%%%%%%%%%%%%%%%%%%%%%%%%%%%%%%%%%%%%%%%%%%%%
%%%%%%%%%%%%%%%%%%%%%%%%%%%%%%%%%%%%%%%%%%%%%%%%%%%%%%%%%%%%%%%%%%%%%%%%%%%%%%%%%%%%%%%%%%%%%%%%%%%%%%%%%

\appendix

\section{The variation of the on-shell 5D action with respect to pseudoscalar sources\label{pseudoscalar2ndvar}}

We consider the scalar sector of the 5D action
(\ref{full5daction}):

\be S_{5D} = \int d^5x \sqrt{g}\Lambda^2 \textrm{tr} \left(
|\partial X|^2+\frac{3}{R^2}|X|^2 \right),\ee see (\ref{5dlag}).

If $X$ is proportional to the unity $\mathbb{I}_{N_f\times N_f}$
matrix, the action in components is rewritten as follows:

\ba S_{5D} = \int d^5x\ N_f\Lambda^2R^3\left[ \frac{\dmu X\dmuu X
- (\dz X)^2}{z^3} + \frac{3}{z^5}X^2 \right]. \ea

The equation of motion for the bulk-to-boundary propagator
$\mathcal{K}(z,x-y)$ is: \be \dz \left(\frac{1}{z^3}\dz
\mathcal{K}(z,x-y)\right) - \frac{1}{z^3}\Box\mathcal{K}(z,x-y) +
\frac{3}{z^5}\mathcal{K}(z,x-y)=0,\ \lim_{z\rightarrow
0}\frac{2}{z}\mathcal{K}(z,x-y)=\delta(x-y.)\ee

The solution in the 4D Euclidean momentum space is \be
\mathcal{K}(z,Q)=Q^2z^2\left(AK_1(Qz)+BI_1(Qz)\right). \ee The
boundary condition implies that $A=\dfrac{1}{2Q}$, and the
$Q\rightarrow 0$ limit -- that $B =
\dfrac{\sigma}{mQ^3}\left(1+\mathcal{O}(1)\right)$, where $\sigma$
is proportional to the chiral condensate $C,
\left<\bar{q}^{\alpha}q^{\beta}\right>\equiv
\Sigma^{\alpha\beta}\equiv C\delta^{\alpha\beta}$. Thus \be
\mathcal{K}(z,Q) = \frac{Qz^2}{2}K_1(Qz)+\frac{\sigma
z^2}{mQ}I_1(Qz).\label{bulktoboundary} \ee

The source of the pseudoscalar current $J^{\alpha\beta}_P(x),
\alpha,\beta=1...N_f$, is related to the boundary value of the $X$
field on the regulatory UV brane: \be
\frac{2}{\epsilon}X^{\alpha\beta}(\epsilon,x)=
iJ^{\alpha\beta}_P(x) + m\delta^{\alpha\beta},
\epsilon\rightarrow0,\ee allowing us to express $X$ in terms of
$J$:

\ba &&X_{\alpha\beta}(z,x)=\int \frac{d^4q}{(2\pi)^4} e^{-iqx}
\mathcal{K}(z,q) \left[iJ^i_P(q)+
\sqrt{2N_f}m\delta(q)\delta^{0i}\right]t^i_{\alpha\beta},\nn
X^{\dag}_{\alpha\beta}(z,x)=\int \frac{d^4q}{(2\pi)^4} e^{-iqx}
\mathcal{K}(z,-q) \left[-iJ^i_P(-q)+
\sqrt{2N_f}m\delta(q)\delta^{0i}\right]t^i_{\alpha\beta},
\label{xbulk}\ea where $t^i_{\alpha\beta}$ are the basis Hermitian
$N_f\times N_f$ matrices, $i=0...N_f^2-1$.

If we denote the integral over $z$ as: \ba &&\mathcal{D}(Q) = \int
dz\left\{
-\frac{\dz\mathcal{K}(z,Q)^2}{z^3}+\left(-\frac{Q^2}{z^3}+\frac{3}{z^5}\right)\mathcal{K}(z,Q)^2
\right\}=\left.\frac{\mathcal{K}(z,Q)\dz\mathcal{K}(z,Q)}{z^3}\right|_{z=\epsilon}\nn
=\frac{1}{4}\left[\frac{1}{\epsilon^2}+\frac{4\sigma}{m}+Q^2\log(Q^2\epsilon^2)\right],
\label{Dapp}\ea the action may be rewritten in terms of the
pseudoscalar sources: \ba
S_{5D}[J_P]&&=\Lambda^2R^3\dfrac{\delta_{ab}}{2}\int
\dfrac{d^4k}{(2\pi)^4} \mathcal{D}(k)\left[iJ^a_P(k)+
\sqrt{2N_f}m\delta(k)\delta^{0a}\right] \left[-iJ^b_P(-k)+
\sqrt{2N_f}m\delta(k)\delta^{0b}\right]\nn+S_{\phi}[J_P],\label{pseudoscalaronshellaction}\ea
where $S_{\phi}[J_P]$ is the action for the longitudinal component
of the axial field $\phi$ which mixes with $X$ \cite{krikun}.

The pseudoscalar two-point correlator is the second variation of
the quadratic action:

\ba &&i\left\langle P_i(q)P_j(p)\right\rangle= \left.-i\
e^{-iS_{5D}[J_P]} \frac{\delta}{\delta
J^i_P(q)}\frac{\delta}{\delta J^j_P(p)}\
e^{iS_{5D}[J_P]}\right|_{J_P=0}\nn =\frac{\delta}{\delta
J^i_P(q)}\frac{\delta}{\delta J^j_P(p)}\
S_{5D}[J_P]=\Lambda^2R^3\delta_{ij}\mathcal{D}(Q)\delta(p+q)+\
\mbox{terms originating from}\ S_{\phi}[J_P].\label{ppapp}\ea

In order to establish the precise connection between $\sigma$ and
the condensate we will calculate the latter as a variation of the
5D action due to a scalar source in the same way as we have
calculated the pseudoscalar correlator \cite{krikun}. In that case
the boundary condition is the following: \be
\frac{2}{\epsilon}X^{\alpha\beta}(\epsilon,x)=
J^{\alpha\beta}_S(x) + m\delta^{\alpha\beta},\ee and the 5D action
is \ba S_{5D}[J_S]=\Lambda^2R^3\dfrac{\delta_{ab}}{2}\int
\dfrac{d^4k}{(2\pi)^4} \mathcal{D}(k)\left[J^a_S(k)+
\sqrt{2N_f}m\delta(k)\delta^{0a}\right] \left[J^b_S(-k)+
\sqrt{2N_f}m\delta(k)\delta^{0b}\right].\label{scalaraction}\ea

Variation of the action yields: \ba
&&\Sigma_{\alpha\beta}(q)=\Sigma^a(q) t^a_{\alpha\beta}=\left.-i
t^a_{\alpha\beta}\ e^{-iS_{5D}[J_S]}\frac{\delta}{\delta
J^a_S(q)}e^{iS_{5D}[J_S]}\right|_{J_S=0}\nn =t^a_{\alpha\beta}
\left.\frac{\delta}{\delta J^a_S(q)}S_{5D}[J_S]\right|_{J_S=0} =
t^a_{\alpha\beta} \Lambda^2R^3\frac{\delta_{ij}}{2}
2\mathcal{D}(k)\delta_{ai}\delta(k-q)\sqrt{2N_f}m\delta(k)\delta_{0j}\nn
=t^0_{\alpha\beta}\sqrt{2N_f} \Lambda^2R^3
m\mathcal{D}(0)\delta(q)=\delta_{\alpha\beta}
\Lambda^2R^3\left(\frac{m}{4\epsilon^2}+\sigma\right)\delta(q).\ea

Thus, \be \Sigma_{\alpha\beta} \equiv C\delta_{\alpha\beta}=
\delta_{\alpha\beta}
\Lambda^2R^3\left(\frac{m}{4\epsilon^2}+\sigma\right)\Rightarrow
C=\Lambda^2R^3\left(\frac{m}{4\epsilon^2}+\sigma\right).\label{sigmaapp}\ee

This allows us to express the pseudoscalar correlator through the
condensate:

\ba i\left\langle
P_i(q)P_j(-q)\right\rangle=\delta_{ij}\left\{\frac{C}{m}+\frac{\Lambda^2R^3}{4}\
Q^2 \log (Q^2\epsilon^2) \right\}+\ \mbox{terms originating from}\
S_{\phi}[J_P]. \ea

%%%%%%%%%%%%%%%%%%%%%%%%%%%%%%%%%%%%%%%%%%%%%%%%%%%%%%%%%%%%%%%%%%%%%%%%%%%%%%%%%%%%%%%%%%%%%%%%%%%%%%%%%
%%%%%%%%%%%%%%%%%%%%%%%%%%%%%%%%%%%%%%%%%%%%%%%%%%%%%%%%%%%%%%%%%%%%%%%%%%%%%%%%%%%%%%%%%%%%%%%%%%%%%%%%%
%%%%%%%%%%%%%%%%%%%%%%%%%%%%%%%%%%%%%%%%%%%%%%%%%%%%%%%%%%%%%%%%%%%%%%%%%%%%%%%%%%%%%%%%%%%%%%%%%%%%%%%%%
%%%%%%%%%%%%%%%%%%%%%%%%%%%%%%%%%%%%%%%%%%%%%%%%%%%%%%%%%%%%%%%%%%%%%%%%%%%%%%%%%%%%%%%%%%%%%%%%%%%%%%%%%

\section{The variation of the on-shell 5D action with respect to scalar sources\label{scalar2ndvar}}

In this appendix we will discuss second derivative of the on-shell
action (\ref{5dlagsoft}) with respect to the mass: \be i\left<
S(0)S(0) \right> = \frac{1}{2}\ \ddmass^2\mathcal{V}\ \frac{\delta
S_{surf}}{\delta X} + \frac{1}{4}\ \ddmass\mathcal{V}\
\frac{\delta^2 S_{surf}}{\delta X^2}\ \ddmass\mathcal{V} +
\frac{1}{4}\ \ddmass\mathcal{V}\ \frac{\delta^2 S_{vol}}{\delta
X^2}\ \ddmass\mathcal{V}.\label{sssoftapp}\ee

The first term here \ba \frac{1}{2}\ \ddmass^2\mathcal{V}\
\frac{\delta S_{surf}}{\delta X} = \frac{1}{2}\left.\left(
\ddmass^2\mathcal{V}(z)\dfrac{e^{-\Phi(z)} \dz}{z^3}\mathcal{V}(z)
+  \mathcal{V}(z)\dfrac{e^{-\Phi(z)}
\dz}{z^3}\ddmass^2\mathcal{V}(z)
\right)\right|_{z=\epsilon}=0,\label{ssterm1}\ea because
$\ddmass^2\mathcal{V}(z) =
-\dfrac{2\sigma^2z^5}{(\gamma-m)^3}\dfrac{\sinh\left(\frac{\sigma
z^2}{\gamma-m}\right)}{\cosh^3\left(\frac{\sigma
z^2}{\gamma-m}\right)} \sim z^7, z\rightarrow 0, \mathcal{V}(z)
\sim z, z\rightarrow 0$.

The second term in (\ref{sssoft}) \be \frac{1}{4}\
\ddmass\mathcal{V}\ \frac{\delta^2 S_{surf}}{\delta X^2}\
\ddmass\mathcal{V} =
\left.\frac{N_f\Lambda^2R^3}{4}\ddmass\mathcal{V}(z)\dfrac{e^{-\Phi(z)}
\dz}{z^3} \ddmass\mathcal{V}(z)\right|_{z=\epsilon} =
N_f\Lambda^2R^3\left( \frac{1}{\epsilon^2}-\frac{1}{4}\kappa m^2
\right).\label{ssterm2}\ee

The third term in (\ref{sssoft}) is the most involved. \ba
&&\frac{1}{4}\ \ddmass\mathcal{V}\ \frac{\delta^2 S_{vol}}{\delta
X^2}\ \ddmass\mathcal{V} = \frac{3}{4}N_f\Lambda^2R^3\kappa
\int_{\epsilon}^{\infty} dz
\frac{e^{-\Phi(z)}}{z^5}\mathcal{V}^2(z)\ddmass\mathcal{V}^2(z)\ns
&= &\frac{3}{8}N_f\Lambda^2R^3\kappa\gamma^2
\int_{\epsilon_x}^{\infty} dx \frac{e^{-\Phi(z(x))}}{x}\left[
\tanh x + \frac{m}{\gamma}(1-\tanh x) \right]^2\left[ 1-\tanh x +
x \cosh^{-2}x \right]^2,\label{ssterm3}\ea where in the last
integral we have introduced a dimensionless variable
$x=\dfrac{\sigma z^2}{\gamma-m}, \epsilon_x=\dfrac{\sigma
\epsilon^2}{\gamma-m}$.

The functions \ba &&f_0(x)=\tanh^2x\left(1-\tanh
x+\frac{x}{\cosh^2x}\right)^2/x,\nn f_1(x)=2\tanh x(1-\tanh
x)\left(1-\tanh x+\frac{x}{\cosh^2x}\right)^2/x,\nn
f_2(x)=(1-\tanh x)^2\left(1-\tanh
x+\frac{x}{\cosh^2x}\right)^2/x,\nonumber\ea decrease rapidly when
$x\gtrsim 1$ and we will use the $z\rightarrow 0$ asymptotic
behavior of the dilaton: $\Phi(z(x)) = \dfrac{\kappa m^2}{4}z^2 =
\dfrac{\kappa m^2 (\gamma-m)}{4\sigma}x \equiv \varkappa x$.

Numerically \be A_0 = \int_0^{\infty} dx\ f_0(x) = 0.377,\ A_1 =
\int_0^{\infty} dx\ f_1(x) = 0.977,\ A_2 = \left.\int_0^{\infty}
dx\ f_2(x)\right|_{reg} = -1.487.\ee

As a result, \ba \frac{1}{4}\ \ddmass\mathcal{V}\ \frac{\delta^2
S_{vol}}{\delta X^2}\ \ddmass\mathcal{V} &=&
\frac{3}{8\pi^2}\lambda N_f N_c A_0 +
\frac{3}{16\pi^2}m\sqrt{\lambda\kappa} N_f N_c A_1 \ns &+&
\frac{3}{32\pi^2}m^2\kappa N_f N_c \left( A_2+\log
\left(\frac{2\pi^2 C
\epsilon^2}{N_c}\sqrt{\frac{\kappa}{\lambda}}\right)-\frac{N_c}{2\pi^2
C}\sqrt{\frac{\lambda^3}{\kappa}} A_0
\right).\label{ssterm3series}\ea

We finally obtain from (\ref{sssoft},\ref{ssterm1},\ref{ssterm2},
\ref{ssterm3},\ref{ssterm3series}): \ba i\left< S_i(0)S_j(0)
\right> &=& \delta_{ij}\left\{ \frac{3}{8\pi^2}\lambda  N_c A_0 +
\frac{3}{16\pi^2}m\sqrt{\lambda\kappa} N_c A_1 +
\frac{N_c}{4\pi^2\epsilon^2}\right.\ns &+&\left.
\frac{3}{32\pi^2}m^2\kappa N_c \left( A_2-\frac{2}{3}+\log
\left(\frac{2\pi^2 C
\epsilon^2}{N_c}\sqrt{\frac{\kappa}{\lambda}}\right)-\frac{N_c}{2\pi^2
C}\sqrt{\frac{\lambda^3}{\kappa}} A_0
\right)\right\}.\label{ssapp} \ea

%%%%%%%%%%%%%%%%%%%%%%%%%%%%%%%%%%%%%%%%%%%%%%%%%%%%%%%%%%%%%%%%%%%%%%%%%%%%%%%%%%%%%%%%%%%%%%%%%%%%%%%%%
%%%%%%%%%%%%%%%%%%%%%%%%%%%%%%%%%%%%%%%%%%%%%%%%%%%%%%%%%%%%%%%%%%%%%%%%%%%%%%%%%%%%%%%%%%%%%%%%%%%%%%%%%
%%%%%%%%%%%%%%%%%%%%%%%%%%%%%%%%%%%%%%%%%%%%%%%%%%%%%%%%%%%%%%%%%%%%%%%%%%%%%%%%%%%%%%%%%%%%%%%%%%%%%%%%%
%%%%%%%%%%%%%%%%%%%%%%%%%%%%%%%%%%%%%%%%%%%%%%%%%%%%%%%%%%%%%%%%%%%%%%%%%%%%%%%%%%%%%%%%%%%%%%%%%%%%%%%%%

\section{The lowest Kaluza--Klein mode of the field $\phi$\label{thephi}}

In this section we will explicitly derive the formula for the
lowest Kaluza--Klein mode $f^{(0)}_{\phi}\equiv f_{\phi}$ of the
field $\phi$ by solving the corresponding equations of motion
perturbatively up to the first order in $m$ and $m^2_{\pi}\propto
m$.

In order to find the first-order corrections to the chiral limit
of the solution $f_{\phi}$ we will solve the following system of
equations \cite{hardwall}, \cite{krikun} of motion perturbatively:
\ba
&&\dz\left(\frac{1}{z}\ \dz\phi\right)+k^2\frac{v^2(z)}{z^3}(\pi-\phi)=0 \label{eom1}\\
&&m^2_{\pi}\dz\phi + k^2\frac{v^2(z)}{z^2}\ \dz\pi=0. \label{eom2}
\ea

Here $k^2=\Lambda^2R^2g_5^2=3,\ \Lambda^2R^3=\dfrac{N_c}{4\pi^2},\
\Lambda^2R^3\sigma=C$. The small parameter of the perturbative
solution is $m$ and $m^2_{\pi}\propto m$. The boundary conditions
are: \be \pi(0)=1,\ \phi(0)=\dz\phi(z_m)=0. \label{bc}\ee We
cannot impose uniform zero boundary conditions on both functions
$\phi$ and $\pi$ without making the equations incompatible.

The solution in the chiral limit is: \ba
\pi^{(0)}(z)&=&const=\pi(0)=1, \label{pi0}\\
\phi^{(0)}(z)&=&1-2\Gamma^{-1}(1/3)\left(\frac{k\sigma}{2}\right)^{1/3}\frac{K_{-2/3}(k\sigma
z^3_m)}{I_{-2/3}(k\sigma z^3_m)}\ zI_{1/3}(k\sigma z^3)\ns
&-&2\Gamma^{-1}(1/3)\left(\frac{k\sigma}{2}\right)^{1/3}zK_{1/3}(k\sigma
z^3).\label{phi0} \ea

We will impose the boundary conditions (\ref{bc}) at $z=0$ and
$z=z_m$ upon the Green function $\mathcal{G}(z,z')$.

The Green function is a solution of the following equation
(\ref{eom1}): \be \left(\dz\frac{1}{z}\
\dz-k^2\sigma^2z^3\right)\mathcal{G}(z,z') = \delta(z-z'),\
\mathcal{G}(0,z')=0, \left.\dz\mathcal{G}(z,z')\right|_{z=z_m},
\label{greeneq}\ee

\ba \mathcal{G}(z,z') &=&-\frac{z z'}{t'}\cdot
\frac{1}{I_{-2/3}(t_m)\left\{ K_{1/3}(t') I_{-2/3}(t') +
K_{-2/3}(t') I_{1/3}(t') \right\}}\ns &\times& \biggl\{
K_{1/3}(t')K_{-2/3}(t_m)I_{1/3}(t) +
\theta(t-t')I_{1/3}(t')I_{-2/3}(t_m)K_{1/3}(t) \ns &+&
\theta(t'-t)K_{1/3}(t')I_{-2/3}(t_m)I_{1/3}(t) \biggl\},\
t=k\sigma z^3,\ t'=k\sigma z'^3,\ t_m=k\sigma
z_m^3.\label{green}\\
&=&-\frac{z z'}{t'}\cdot \frac{1}{I_{-2/3}(t_m)\left\{ K_{1/3}(t')
I_{-2/3}(t') + K_{-2/3}(t') I_{1/3}(t') \right\}}\ns &\times&
\biggl\{
K_{1/3}(t')\biggl(K_{-2/3}(t_m)+I_{-2/3}(t_m)\biggr)I_{1/3}(t)\ns
&+& \theta(t-t')I_{-2/3}(t_m)\biggl(I_{1/3}(t')K_{1/3}(t) -
K_{1/3}(t')I_{1/3}(t)\biggr) \biggl\}\label{green1}\\
&\equiv& \mathcal{F}_0(z')\ zI_{1/3}(k\sigma z^3) +
\theta(z-z')\biggl( \mathcal{F}_I(z')\ zI_{1/3}(k\sigma z^3) +
\mathcal{F}_K(z')\ zK_{1/3}(k\sigma z^3) \biggr).\label{green2}\ea

The first-order correction to $\pi(z)$ is (\ref{eom2}): \be
\pi^{(1)}(z)=-\frac{m^2_{\pi}}{k^2}\int_0^z du
\frac{u^2\du\phi^{(0)}(u)}{v^2(u)},\label{pi1} \ee and the first
order correction to $\phi(z)$ is (\ref{eom1}): \be
\phi^{(1)}(z)=\int_0^{z_m} dw\ \mathcal{G}(z,w)\biggl(2m\sigma
k^2w\left(\phi^{(0)}(w)-\pi^{(0)}(w)\right)-k^2\sigma^2w^3\pi^{(1)}(w)\biggr).\label{phi1}\ee

We have retained the terms $\propto m,\ m^2$ in the $v^2(u)$
function in the expression (\ref{pi1}) for the sake of convergence
of the integral although we exceed the necessary accuracy level.

Substituting (\ref{phi0},\ \ref{pi0},\ \ref{pi1}) into
(\ref{phi1}) and using the form (\ref{green2}) we obtain: \ba
\phi^{(1)}(z) &=&
-2\Gamma^{-1}(1/3)\left(\frac{k\sigma}{2}\right)^{1/3}
\int_0^{z_m} dw\ \mathcal{G}(z,w)\left\{2m\sigma k^2w^2 \left(
I_{1/3}(k\sigma w^3)\frac{K_{-2/3}(k\sigma
z^3_m)}{I_{-2/3}(k\sigma z^3_m)}\right.\right. \ns &+&
K_{1/3}(k\sigma w^3) \biggr) + 3k\sigma^3 m^2_{\pi}w^3 \int_0^w
du\ \frac{u^5}{v^2(u)}\ns &\times&\left.\left( I_{-2/3}(k\sigma
u^3)\frac{K_{-2/3}(k\sigma z^3_m)}{I_{-2/3}(k\sigma z^3_m)} -
K_{-2/3}(k\sigma u^3) \right) \right\}\label{phi1full1}\\
&=& \mu_0\cdot zI_{1/3}(k\sigma z^3) + \mu_I(z)\cdot
zI_{1/3}(k\sigma z^3) + \mu_K(z)\cdot zK_{1/3}(k\sigma
z^3).\label{phi1full2} \ea

Here we have introduced the following notations (\ref{green1},\
\ref{green2}):

\ba \mu_0 &=& \int_0^{z_m} dw\ \mathcal{F}_0(w)\biggl( 2m\sigma
k^2w\left( \phi^{(0)}(w)-\pi^{(0)}(w) \right)
-k^2\sigma^2w^3\pi^{(1)}(w) \biggr),\label{mu0}\\
\mu_I(z) &=& \int_0^{z} dw\ \mathcal{F}_I(w)\biggl( 2m\sigma
k^2w\left( \phi^{(0)}(w)-\pi^{(0)}(w) \right)
-k^2\sigma^2w^3\pi^{(1)}(w) \biggr),\label{muI}\\
\mu_K(z) &=& \int_0^{z} dw\ \mathcal{F}_K(w)\biggl( 2m\sigma
k^2w\left( \phi^{(0)}(w)-\pi^{(0)}(w) \right)
-k^2\sigma^2w^3\pi^{(1)}(w) \biggr).\label{muK} \ea

In terms of these definitions we can express the solution $\phi$
(\ref{phi0},\ \ref{phi1},\ \ref{phi1full1},\ \ref{phi1full2}):

\ba \phi(z) &=&
1-2\Gamma^{-1}(1/3)\left(\frac{k\sigma}{2}\right)^{1/3}\frac{K_{-2/3}(k\sigma
z^3_m)}{I_{-2/3}(k\sigma z^3_m)}\ zI_{1/3}(k\sigma z^3)
-2\Gamma^{-1}(1/3)\left(\frac{k\sigma}{2}\right)^{1/3}zK_{1/3}(k\sigma
z^3)\ns &+& \mu_0 zI_{1/3}(k\sigma z^3) + \mu_I(z)zI_{1/3}(k\sigma
z^3) + \mu_K(z)zK_{1/3}(k\sigma z^3) +
\mathcal{O}(m^2).\label{phitot}\ea

%%%%%%%%%%%%%%%%%%%%%%%%%%%%%%%%%%%%%%%%%%%%%%%%%%%%%%%%%%%%%%%%%%%%%%%%%%%%%%%%%%%%%%%%%%%%%%%%%%%%%%%%%
%%%%%%%%%%%%%%%%%%%%%%%%%%%%%%%%%%%%%%%%%%%%%%%%%%%%%%%%%%%%%%%%%%%%%%%%%%%%%%%%%%%%%%%%%%%%%%%%%%%%%%%%%
%%%%%%%%%%%%%%%%%%%%%%%%%%%%%%%%%%%%%%%%%%%%%%%%%%%%%%%%%%%%%%%%%%%%%%%%%%%%%%%%%%%%%%%%%%%%%%%%%%%%%%%%%
%%%%%%%%%%%%%%%%%%%%%%%%%%%%%%%%%%%%%%%%%%%%%%%%%%%%%%%%%%%%%%%%%%%%%%%%%%%%%%%%%%%%%%%%%%%%%%%%%%%%%%%%%

\section{The canonical normalization of the $\phi$ field and the parameter $L_4$ of the chiral Lagrangian\label{chiralparnorm}}

In this section we will present the explicit expressions for the
quantities $N_{\pi}$ and $g_{m\ \phi^2}$ introduced in the section
{\bf \ref{chirallag}} eq. (\ref{chiralfromads}).

Let us consider the quadratic part of the 5D action
(\ref{full5daction}) where we only take into account the first
Kaluza--Klein mode of the field $\phi$ which corresponds to the
pion. We obtain:

\be S_{5D} \rightarrow \int dz\left(
\Lambda^2R^3\frac{v^2(z)}{z^3}f^2_{\phi}(z)+\frac{R}{g^2_5}\frac{1}{z}\dz
f^2_{\phi}(z) \right)\times \int d^4x\ \frac{1}{2}\ \dmu
\phi^a(x)\dmuu \phi^a(x).\label{5dlagphi} \ee

Following (\ref{chiralfromads}) let us denote
\ba\lim_{m\rightarrow 0} \int dz\left(
\Lambda^2R^3\dfrac{v^2(z)}{z^3}f^2_{\phi}(z)+\dfrac{R}{g^2_5}\dfrac{1}{z}\dz
f^2_{\phi}(z) \right) &=& \int dz\left( \Lambda^2R^3\sigma^2
z^3\left.f^2_{\phi}(z)\right|_{m=0} \right. \ns &+&\left.
\dfrac{R}{g^2_5}\dfrac{1}{z}\dz \left.f^2_{\phi}(z)\right|_{m=0}
\right) = N^2_{\pi}\label{Npi}\ea. We immediately obtain: \be
F_{\pi}\pi^a = N_{\pi}\phi^a, \ee where
$\left.f^2_{\phi}(z)\right|_{m=0}$ may be extracted from
(\ref{phitot},\ref{phimode}).

The term of the Chiral Lagrangian \ba \mathcal{P}_4 =
\mathrm{tr}(D_{\mu}U^{\dag}D^{\mu}U)\mathrm{tr}(U^{\dag}\chi+\chi^{\dag}U)=
8\frac{\Sigma}{F^2_{\pi}}m \dmu \pi^a(x)\dmuu \pi^a(x) +
\mathcal{O}(\pi^4)\label{p4}\ea from the AdS/QCD point of view is
generated by the action (\ref{5dlagphi}): \ba S_{5D} &\rightarrow&
m \int d^4x\ \dmu \phi^a(x)\dmuu \phi^a(x)\times \int dz\left(
\frac{1}{2}\Lambda^2R^3\sigma z \left.f^2_{\phi}(z)\right|_{m=0}
\right.\ns &+&\left. \Lambda^2R^3\sigma^2 z^3 \left.f_{\phi}(z)
\ddmass f_{\phi}(z)\right|_{m=0} + \frac{R}{g^2_5 z}\left.\dz
f_{\phi}(z) \ddmass \dz f_{\phi}(z)\right|_{m=0}
 \right) .\ea
If we denote according to (\ref{chiralfromads})

\ba &&\int dz\left( \frac{1}{2}\Lambda^2R^3\sigma z
\left.f^2_{\phi}(z)\right|_{m=0} + \Lambda^2R^3\sigma^2 z^3
\left.f_{\phi}(z) \ddmass f_{\phi}(z)\right|_{m=0} \right.\ns &+&
\left. \frac{R}{g^2_5 z}\left.\dz f_{\phi}(z) \ddmass \dz
f_{\phi}(z)\right|_{m=0}
 \right) = g_{m\ \phi^2},\label{gmphi2}\ea

 we obtain:
 \be L_4=\frac{F^4_{\pi}}{8\Sigma}g_{m\ \phi^2} N^{-2}_{\pi},\label{appl4}\ee
where $\left.f^2_{\phi}(z)\right|_{m=0}$ and $\left.\ddmass
f^2_{\phi}(z)\right|_{m=0}$ may be extracted from
(\ref{green}--\ref{phitot},\ \ref{phimode}).

%%%%%%%%%%%%%%%%%%%%%%%%%%%%%%%%%%%%%%%%%%%%%%%%%%%%%%%%%%%%%%%%%%%%%%%%%%%%%%%%%%%%%%%%%%%%%%%%%%%%%%%%%
%%%%%%%%%%%%%%%%%%%%%%%%%%%%%%%%%%%%%%%%%%%%%%%%%%%%%%%%%%%%%%%%%%%%%%%%%%%%%%%%%%%%%%%%%%%%%%%%%%%%%%%%%
%%%%%%%%%%%%%%%%%%%%%%%%%%%%%%%%%%%%%%%%%%%%%%%%%%%%%%%%%%%%%%%%%%%%%%%%%%%%%%%%%%%%%%%%%%%%%%%%%%%%%%%%%
%%%%%%%%%%%%%%%%%%%%%%%%%%%%%%%%%%%%%%%%%%%%%%%%%%%%%%%%%%%%%%%%%%%%%%%%%%%%%%%%%%%%%%%%%%%%%%%%%%%%%%%%%

\section{The parameters $L_1, L_2$ and $L_3$ of the chiral Lagrangian\label{chiralpar}}

In this section we will present the explicit expression for the
quantity $g_{\phi^4}$ introduced in the section {\bf
\ref{chirallag}} eq. (\ref{chiralfromads}).

The $\phi^4$ interaction (the Skyrme--like term in
(\ref{chiralfromads})) is generated by the quartic part of the
gauge sector of the full 5D action (\ref{gaugeaction}): \be
S_{\phi^4}=\int d^5x
\frac{-R}{4g_5^2z}f^{abe}f^{cde}\dmu\phi^a\dnu\phi^b\dmuu\phi^c\dnuu\phi^d.
\label{sphi4}\ee

The solution (\ref{phitot}) for $\phi(z)$ from subsection {\bf
\ref{thephi}} is the function $f_{\phi}(z)$ with which we were
dealing in section {\bf \ref{chirallag}}, eq. (\ref{kkmodes}) . We
will restrict our consideration to the chiral limit. This
corresponds to the zero-order solution $\phi^{(0)}(z)$
(\ref{phi0}): \ba f_{\phi}(z) &=& N_{\phi}\biggl(
1-2\Gamma^{-1}(1/3)\left(\frac{k\sigma}{2}\right)^{1/3}\frac{K_{-2/3}(k\sigma
z^3_m)}{I_{-2/3}(k\sigma z^3_m)}\ zI_{1/3}(k\sigma z^3)\ns &-&
2\Gamma^{-1}(1/3)\left(\frac{k\sigma}{2}\right)^{1/3}zK_{1/3}(k\sigma
z^3) \biggr).\label{phimode} \ea

The normalization factor $N_{\phi}$ is determined, as in
(\ref{vmode}), by the condition (\ref{norm}): \be
\int_{\epsilon}^{z_m} \frac{dz}{z}\ f^2_{\phi}(z) =
1.\label{normphi}\ee

The effective 4D coupling is (\ref{sphi4},\ \ref{kkmodes}): \ba
g_{\phi^4}&=& \int dz \frac{-R}{4g_5^2z}f^4_{\phi}(z)=
N^4_{\phi}\int_{\epsilon}^{z_m} dz \frac{-R}{4g_5^2z}
\times\biggl( 1
-2\Gamma^{-1}(1/3)\left(\frac{k\sigma}{2}\right)^{1/3} \ns
&\times&\frac{K_{-2/3}(k\sigma z^3_m)}{I_{-2/3}(k\sigma z^3_m)}\
zI_{1/3}(k\sigma z^3)
-2\Gamma^{-1}(1/3)\left(\frac{k\sigma}{2}\right)^{1/3}zK_{1/3}(k\sigma
z^3) \biggr)^4. \label{gphi4}\ea

We obtain an effective 4D Lagrangian for the pions: \ba
\mathcal{L}_{\pi^4} =
g_{\phi^4}\frac{F^4_{\pi}}{N^4_{\pi}}f^{abe}f^{cde}\dmu \pi^a \dnu
\pi^b \dmuu \pi^c \dnuu \pi^d ,\label{4pionlag}\ea where $N_{\pi}$
is the normalization factor that determines the proportionality
between the fields $\pi^a$ and $\phi^a$, see the subsection {\bf
\ref{chiralparnorm}}, eq. (\ref{Npi}).

%%%%%%%%%%%%%%%%%%%%%%%%%%%%%%%%%%%%%%%%%%%%%%%%%%%%%%%%%%%%%%%%%%%%%%%%%%%%%%%%%%%%%%%%%%%%%%%%%%%%%%%%%
%%%%%%%%%%%%%%%%%%%%%%%%%%%%%%%%%%%%%%%%%%%%%%%%%%%%%%%%%%%%%%%%%%%%%%%%%%%%%%%%%%%%%%%%%%%%%%%%%%%%%%%%%
%%%%%%%%%%%%%%%%%%%%%%%%%%%%%%%%%%%%%%%%%%%%%%%%%%%%%%%%%%%%%%%%%%%%%%%%%%%%%%%%%%%%%%%%%%%%%%%%%%%%%%%%%
%%%%%%%%%%%%%%%%%%%%%%%%%%%%%%%%%%%%%%%%%%%%%%%%%%%%%%%%%%%%%%%%%%%%%%%%%%%%%%%%%%%%%%%%%%%%%%%%%%%%%%%%%

\end{document}